\documentclass[aip,reprint]{revtex4-1}

\usepackage{amsmath}
\usepackage{amssymb}
\usepackage{graphicx}
\usepackage{dcolumn}
\usepackage{bm}
\usepackage{multirow}
\usepackage{hyperref}
\usepackage{color}
\usepackage{xcolor}

\usepackage{appendix}

\newcommand{\newtext}[1]{\textcolor{black}{#1}}

\begin{document}

\title{Symmetry-specific characterization of bond orientation order in DNA-assembled nanoparticle lattices}

\author{Jack A. Logan}
\affiliation{Department of Mechanical Engineering \& Materials Science, Yale University, New Haven, Connecticut 06520, USA}
\author{Aaron Michelson}
\affiliation{Department of Chemical Engineering, Columbia
University, 817 SW Mudd, New York, NY 10027,
USA}
\affiliation{Center for Functional Nanomaterials, Brookhaven National Laboratory, Upton, NY 11973, USA}
\author{Ajith Pattammattel}
\author{Hanfei Yan}
\affiliation{National Synchrotron Light Source II, Brookhaven National Laboratory, Upton, New York 11973, USA}
\author{Oleg Gang}
\affiliation{National Synchrotron Light Source II, Brookhaven National Laboratory, Upton, New York 11973, USA}
\affiliation{Department of Chemical Engineering, Columbia
University, 817 SW Mudd, New York, NY 10027,
USA}
\affiliation{Center for Functional Nanomaterials, Brookhaven National Laboratory, Upton, NY 11973, USA}
\author{Alexei V. Tkachenko}
\email{oleksiyt@bnl.gov}
\affiliation{Center for Functional Nanomaterials, Brookhaven National Laboratory, Upton, NY 11973, USA}

\begin{abstract}
 Bond-orientational order in DNA-assembled nanoparticles lattices is explored with the help of recently introduced  Symmetry-specific Bond Order Parameters (SymBOPs).  This approach provides a more sensitive analysis of local order than traditional scalar Bond Order Parameters, facilitating the identification of coherent domains at the single bond level. The present study expands the method initially developed for assemblies of  anisotropic particles to the isotropic ones or cases where particle orientation information is unavailable. The SymBOP analysis was applied  to experiments on  DNA-frame-based assembly of nanoparticle lattices. It proved highly sensitive in identifying coherent crystalline domains with different orientations, as well as  detecting  topological defects, such as dislocations.  Furthermore, the analysis  distinguishes individual sublattices within a single crystalline domain, such as  pair of interpenetrating FCC lattices within a cubic diamond. The results underscore the versatility and robustness of SymBOPs in characterizing ordering phenomena, making them valuable tools for investigating structural properties in various systems.

\end{abstract}


\maketitle

\section{\label{sec:Introduction} Introduction}

The concept of bond-orientational order is crucial for understanding crystals \cite{Frenkel1996, Frenkel2004, BOP2008,BOP2018,nguyen2015identification,keys2011characterizing,santiso2011general}, quasi-crystals \cite{quasi}, glasses \cite{Steinhardt, tanaka1999,tanaka2012, Vasilyev_1, Vasilyev_chrom}, and even morphodenesis in living systems \cite{cells}. Bond Order Parameters (BOPs) have been developed to quantify this order, initially in 2D crystallization \cite{Nelson_2D,Nelson_PRB},  and later extended to 3D  structures\cite{Hess1980,MITUS_81,MITUS_82, Nelson_Toner}. 
 Steinhardt et al.  in their classical paper \cite{Steinhardt} introduced rotationally invariant BOPs, such as $Q_l$ and $W_l$. These scalars are widely used as descriptors for the degree of crystallinity, for classifying particles as ``crystal-like" or ``amorphous-like", as well as for  identifying structural motifs like FCC and BCC \cite{Frenkel1996,Frenkel2004,BOP2008,BOP2018}. However, they lack important features of classical order parameters, as they do not vanish in the disordered case and do not capture the orientation of ordered structures.

To address these limitations, we have recently introduced Symmetry-specific Bond Order Parameters (SymBOPs) \cite{logan2022symmetry,mushnoori2022controlling},  invariant under specific point symmetries. SymBOPs  provide a more sensitive characterization of the local order compared to traditional BOPs, enable a meaningful structural characterization down to the single bond level, and facilitate the resolution of individual coherent domains. Our initial focus was on simulations of anisotropic building blocks, inspired by the progress in the self-assembly of sophisticated ``designer particles" \cite{Glotzer_Solomon2007,Pine_patchy_2003,Oleg_hybrid_2016,diamond_oleg,Scior_Nat_phys, sun2020valence, michelson2022three}. In that setting, the particle anisotropy dictates a natural choice of the  reference coordinate system needed for an  unambiguous  definition of SymBOPs.

In this paper, we present the application of SymBOPs to assemblies of anisotropic particles and cases where information about particle orientations is unavailable. Extending the analysis beyond simulated structures, it is also applied to real experimental data on DNA-based nanoscale self-assembly \cite{diamond_oleg,michelson2022three}. We demonstrate the effectiveness of SymBOPs as a characterization tool capable of identifying coherent domains and structural defects in both simulated and experimental systems.

\section{\label{sec:Orientational Order} Symmetry-specific bond  order parameters}
\subsection{SymBOP definition}

In this work, we use the term ``bond" in a geometric rather than physical sense: two particles are defined to be neighbors if the separation between them is below a certain threshold. Let $i$ and $j$ be two neighboring particles whose relative position is ${\bf r}_{ij} = {\bf r}_j - {\bf r}_i$. The orientation of the bond between them is given by a unit  vector  ${\bf \hat{b}}_{ij} = {\bf r}_{ij}/r_{ij}$. As discussed in Ref. \citenum{logan2022symmetry},  the bond orientational order is defined in terms of globally averaged bond multipoles. The latter are symmetric $l$-th rank tensors constructed from bond vectors ${\bf \hat{b}}_{i j}$. In  2D,  it may be equivalently represented by a complex Bond Order Parameter (BOP):
\begin{equation}
\label{BOP2D} 
\Psi_l=\langle e^{{\rm i}l\psi(\hat{b}_{i j})}\rangle 
\end{equation}
Here $\psi(\hat{b}_{i j})$ is the polar angle associated with the bond direction. 

In the 3D case, a convenient representation of the bond multipole is provided by  the spherical harmonics: 
\begin{equation}
     |Q)_l = \left\langle |{\bf \hat{b}}_{ij})_l\right\rangle 
\end{equation}
Here the bra-ket notation   $|\ldots)_l$  represents  the $l$-th degree spherical harmonics that form  a $(2l+1)$-dimensional complex-valued  vector  \cite{logan2022symmetry}. In particular, a single-bond multipole is 
\begin{equation}
|{\bf \hat{b}}_{ij})_l = \left\{ \sqrt{\frac{4\pi}{2l+1}} Y_{l}^m({\bf \hat{b}}_{ij}) \right\}_{m=-l,\ldots,l}
\end{equation}
The system-averaged bond multipole $|Q)_l$ generates a number of rotational invariants that are widely used as scalar BOPs. Those include  $ Q_l=\sqrt{\left(Q|Q\right)_l}$, as well as the higher-order invariants $W_l$ \cite{Steinhardt,Frenkel1996, Frenkel2004}. These scalars are useful descriptors  of the structure, but they don't contain all the important structural information encoded by the multipole $|Q)_l$. This can be contrasted with  the  2D case, Eq (\ref{BOP2D}), where a non-zero BOP $\Psi_l$ indicates the breaking of full rotational symmetry and the emergence of $l$-fold symmetry instead. In addition, the phase of that complex order parameter encodes the orientation of the ordered domain. Below we demonstrate how a similar family of order parameters can be constructed in 3D.  

Consider a case when the overall 3D structure possesses certain point symmetry. This implies that $|Q)_l$ must belong to a subspace invariant under this symmetry transformation, i.e., in the thermodynamic limit 
\begin{equation}
\label{eq:SymBOP}
    |Q)_l\rightarrow \widehat{\mathcal{P}}_g  |Q)_l\equiv |Q^*)_l
\end{equation}
Here $\widehat{\mathcal{P}}_g $ is the projection operator associated with the symmetry group $g$. The right-hand side of Eq. (\ref{eq:SymBOP}) , $|Q^*)_l\equiv\widehat{\mathcal{P}}_g  |Q)_l$ is the Symmetrized Bond Order Parameter (SymBOP) introduced in our recent works \cite{logan2022symmetry}. While it does  coincide with multipole $|Q)_l$ in the ideal case of an infinite uniform crystal, the symmetrization by the  projection operator $\widehat{\mathcal{P}}_g$ provides a valuable refinement for the characterization of realistic systems containing multiple finite domains. 

One of the key applications of the BOPs is the characterization of the local order. In particular, traditional scalar BOPs  are commonly used to identify ``crystal"-like and ``liquid"-like particles in simulated structures. SymBOPs  may distinguish different domains of similarly ordered structures and   may be applied even on the single-bond level. Specifically, the local SymBOP is defined for a single bond between particles $i$ and $j$  as:
\begin{equation}\label{eq:bond_level}
    |{\bf b}^*_{ij})_l = \mathcal{P}_g |\mathbf{\hat{b}}_{ij})_l
\end{equation}
This parameter effectively introduces an equivalence relationship between various bonds that might not have the same orientation  but are related by the symmetry transformation. 

By its definition, a SymBOP depends on the underlying point symmetry group $g$. In 3D, this typically implies the choice of at least one preferential direction (or several equivalent ones). There are two fundamental approaches that one can employ to construct the projection operator $\widehat{\mathcal{P}}_g$, and thus the corresponding SymBOP:
\begin{itemize}
\item{\it Particle reference} As we have demonstrated in earlier works \cite{}, the SymBOP is natural for the characterization of self-assembly of  anisotropic particles. In that case, the system has two types of orientational order: (i) the one  associated with particle orientations and characterized by a polyhedral nematic order parameter, and (ii) the bond orientational order. Thus, the SymBOP can be constructed with respect to the preferred axes of the polyhedral nematic. For instance, if $|S)_l$ is the underlying cubatic order parameter \cite{cubatic_tensor} (in spherical harmonic representation), the corresponding projection operator of the cubic symmetry group is $\widehat{\mathcal{P}}_c=\frac{|S)_l(S|}{(S|S)_l}$.   However, this convenience is lost for isotropic particles or for experimental data where the precise orientation of the building blocks (if originating from anisotropic, e.g., DNA frames, etc.) cannot be determined.
\item {\it Self-consistent reference.} In this case, the  coordinate system  $(\hat{\bf x},\hat{\bf y},\hat{\bf z})$, associated with the projection operator   $\widehat{\mathcal{P}}_g$, is rotated  to maximize the  magnitude of the SymBOP averaged over a certain part of the system. On the other hand, that part itself is determined by the selection of bonds maximally aligned with the ideal structure.   Specifically, by using the bond-level SymBOP, Eq.(\ref{eq:bond_level}), one can identify bonds that belong to the same equivalence class, and then employ the bond percolation procedure to identify the entire coherent domain. This is the approach that we will use in this work.   
\end{itemize}

\subsection{SymBOPs for tetragonal symmetry}
Consider for example a tetragonal crystalline structure that is invariant with respect to $z\rightarrow -z$ reflection and 4-fold rotation about $z$ axis.  The only spherical harmonics invariant under these symmetry transformations are those with $m$ divisible by 4 and even $l$.  In particular, for $l=4$ and $6$, the invariant subspace is three-dimensional, and its  basis vectors are $|{\hat e}^m)_l$, $m=-4,0,4$. This allows one to construct the projection operator $\widehat{\mathcal{P}}_t $ for the case of tetragonal symmetry:
\begin{equation}\label{eq:P_tetragon}
    \mathcal{P}_t^{mm'} (\hat{z})=\begin{cases} 1 & \text{ if}\  m=m'\in\{-4,0,4\}\\
    0 & \text{ otherwise}
    \end{cases}
\end{equation}
Thus, the only non-vanishing components of the tetragonally-symmetric multipole  $|Q^*)_l$ are $Q_l^4$, $Q_l^0$, and $Q_l^{-4}=\overline{Q}_l^{4}$. The  general form of the SymBOP with tetragonal symmetry (for $l=4,6)$ is therefore
\begin{equation}\label{eq:tetragon1}
    |Q^*)_l=Q_l^4(\hat \nu)|{\hat e}^4(\hat \nu))_l+Q_l^0(\hat \nu)|{\hat e}^0(\hat \nu))_l+\overline{Q}_l^4(\hat \nu)|{\hat e}^{-4}(\hat \nu))_l
\end{equation}
Here we have generalized the above result by rotating the coordinate system  to align $\hat z$ with the direction of the main symmetry axis of the tetragonal structure, $\hat \nu$. Each term $\left|{\hat e}^{m}(\hat \nu)\right)_l$  represents a  standard basis vector for the spherical harmonics in this rotated coordinate system.  Note that  the general rotation is parameterized by three Euler angles, $(\theta,\phi,\psi)$, but we only utilize two of them, $\theta$ and $\phi$ to align $\hat z$ with the symmetry direction $\hat \nu$.  The third angle, $\psi$, corresponds to an axial rotation  about the $\hat \nu$ direction.  The component  $Q_l^4$ characterizes the  4-fold symmetry about that axis, as well as the axial orientation of the structure. In fact, this complex parameter can be used as a reduced version of the full tetragonal SymBOP:
\begin{equation}\label{eq:tetragon}
    Q_l^4(\hat \nu)=\left(Q|{\hat e}^4(\hat \nu)\right)_l
\end{equation}
This is a direct analog of the complex BOP that characterizes $l$-fold symmetry  in 2D, Eq. (\ref{BOP2D}). In particular, its non-zero value indicates the breaking of the rotational symmetry and its phase encodes the axial orientation of the ordered domain. For this reason,  we will refer to this single complex component as the tetragonal  SymBOP. The remaining real-valued component $Q_l^0(\hat \nu)$ of the ``full" SymBOP given by Eq. (\ref{eq:tetragon1}), may give additional information about the structure, but, in general, does not characterize its symmetry. An important exception  to this is the cubic symmetry discussed below. 

\subsection{SymBOPs for cubic  symmetry}
Our next example is a crystalline structure with  cubic symmetry. It can be seen as a special case of a tetragonal lattice, for which the $z$ direction  becomes equivalent to $x$ and $y$. This additional symmetry  imposes a constraint on the ratio  $Q_l^0/\left| Q_l^4 \right|$.  Specifically, its values are $\sqrt{14/5}$ and   $-\sqrt{2/7}$ for $l=4$ and $l=6$, respectively. The corresponding  general form of the bond multipole with cubic symmetry is $Q_l\left|\hat{R}_c(\hat{\nu},\psi)\right)_l$, where   $Q_l$ is the traditional scalar BOP that depends on a specific lattice type, e.g., $Q_4\approx 0.19$ and $Q_6 \approx 0.57$ for FCC. Aside from that prefactor, the  cubic symmetry predetermined the form of the bond multipole for $l=4,6$:
\begin{align}
 \left|\hat{R}_c(\hat{\nu},\psi)\right)_l &= \chi_l\left |{\hat e}^0(\hat \nu)\right)_l+  \label{eq:ref_vector_cubic_symmetry} \\
    \nonumber    +&\sqrt{\frac{1-\chi_l^2}{2}}\left[\left|{\hat e}^4(\hat{\nu},\psi)\right)_l+\left|{\hat e}^{-4}(\hat{\nu},\psi)\right)_l \right]
\end{align}
Here $\chi_4=\sqrt{7/12}$  and  $\chi_6=-\sqrt{2}/4$.  $\psi$ is the axial  rotation angle about the $\hat{\nu}$ direction, and  $\left|{\hat e}^{m}(\hat{\nu},\psi)\right)_l\equiv \mathrm{e}^{\rm{i}m\psi}\left|{\hat e}^{m}(\hat{\nu})\right)_l$. Below, we will refer to  $\left|\hat{R}_c(\hat{\nu},\psi)\right)_l$ as {\it reference vector} that characterizes particular orientation of a crystalline domain with underlying cubic symmetry. Note that it is a vector in the spherical harmonic space. From the point of view of the real space, it is in fact  the $l$-th order tensor \cite{logan2022symmetry}. 

The projection operator for the cubic symmetry group is given by  
\begin{equation}
    \widehat{\mathcal{P}}_c (\hat{\nu},\psi)= \left|\hat{R}_c(\hat{\nu},\psi)\right)_l\left(\hat{R}_c(\hat{\nu},\psi)\right|_l
\end{equation} 
Similarly to the tetragonal case, the projection operator  explicitly depends on the choice of the direction $\hat{\nu}$. However, now one needs to additionally specify the axial rotation angle $\psi$. This is because the rotation about  the $\hat \nu$ direction commutes with the symmetries of the tetragonal group, but not with those of the cubic one.

By our definition, the cubic SymBOP is $\widehat{\mathcal{P}}_c (\hat{\nu},\psi)|Q)_l$. The degree of cubic ordering for a specific orientation can be characterized  by its magnitude:
\begin{align}   
\label{cubic_SymBOP}
\left|Q^*_l(\hat{\nu},\psi)\right|&=\left|\widehat{\mathcal{P}}_c (\hat{\nu},\psi)|Q)_l\right|=\left|\left(Q|\hat{R}_c(\hat{\nu},\psi)\right)_l\right|=\\
\nonumber =&\left||Q_l^4(\hat \nu)|\sqrt{2(1-\chi_l^2)}\cos(4(\psi-\psi_0))+\chi_lQ_l^0(\hat \nu)\right|
\end{align}
 Here $Q_l^4(\hat \nu)$ and $Q_l^0(\hat \nu)$ are the components of the global bond monopole $|Q)_l$ rotated in such a way that $\hat z$ coincides with the direction $\hat \nu$. This rotation involves two Euler angles: $\theta$ and $\phi$. The phase factor $\psi_0$ determines the third (axial) rotation angle $\psi$  needed to fully align the global coordinate system $(x,y,z)$ with the symmetry axes of the given cubic structure.

\section{\label{sec:Use of SymBOPs} Self-consistent  domain detection}
In this section, we develop the  SymBOP-based methodology for identifying coherent crystalline domains and 
demonstrate its application  both to simulated and experimental data. The outline of the self-consistent  procedure is as follows: 
\begin{itemize}
\item{As the first step, we find the orientation of the coordinate system that maximizes  the appropriate SymBOP signal, i.e., $|Q^*_l|$. This yields the initial approximation to the orientations of the symmetry axes of the largest ordered domain.}
\item{Using bond-level SymBOPs as classifiers, we identify the groups of bonds that potentially belong to the above domain.}
\item{Finally, by using the bond percolation procedure we find  the entire domain. In the process, the orientation of  the preferred coordinate system is refined to maximize the SymBOP signal for the identified domain. Once the procedure is completed, the bonds that belong to the largest domains are excluded from the further analysis, and we return to  Step 1 to find the signature of the next ordered domain. }
\end{itemize}

\subsection{Application to simulation data}

\begin{figure}[htb!]
       \centering
  \begin{flushleft}{A} \end{flushleft} 
\includegraphics[width=0.9\linewidth]{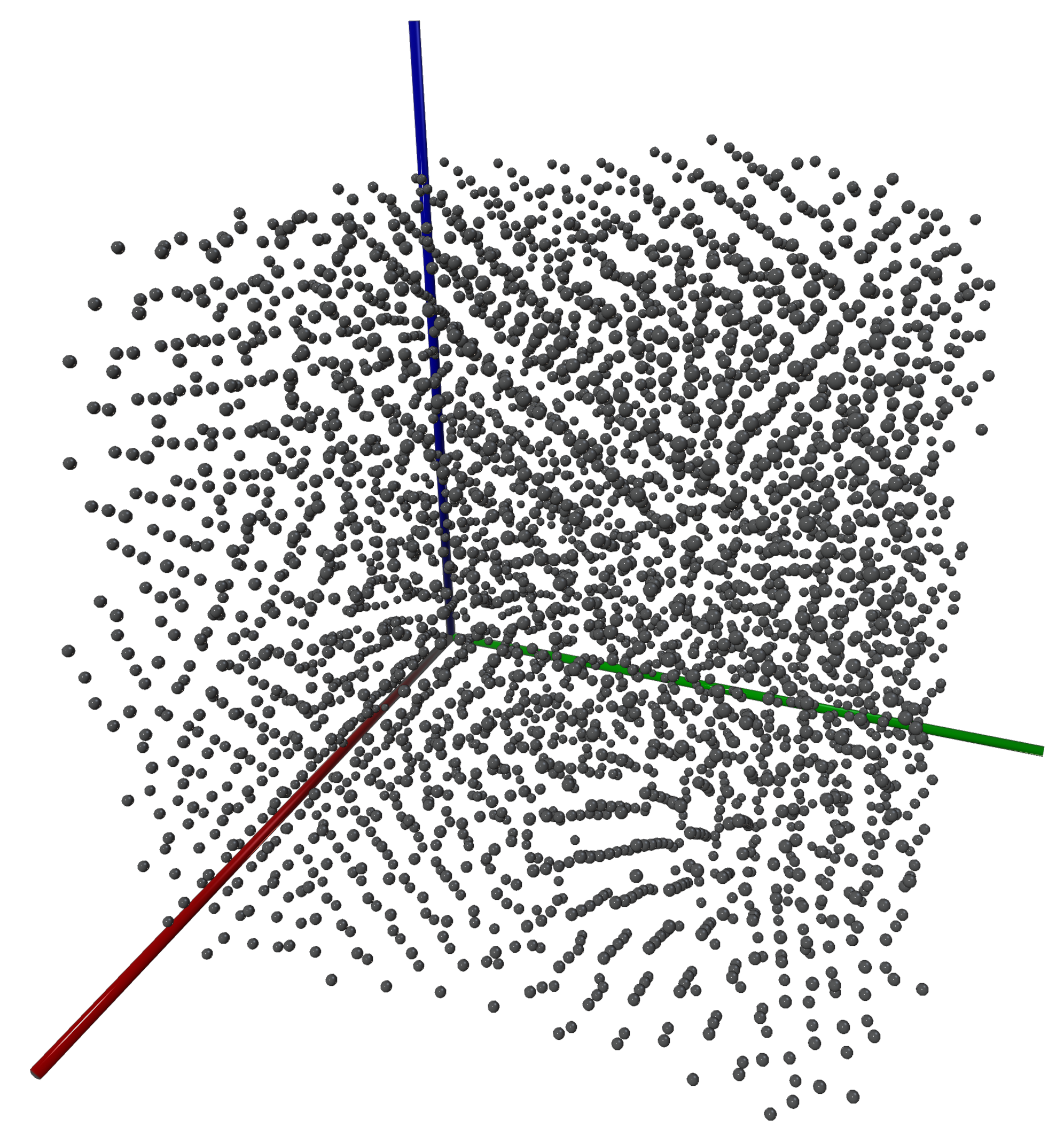}
\begin{flushleft}{B} \end{flushleft} 
\includegraphics[width=\linewidth]{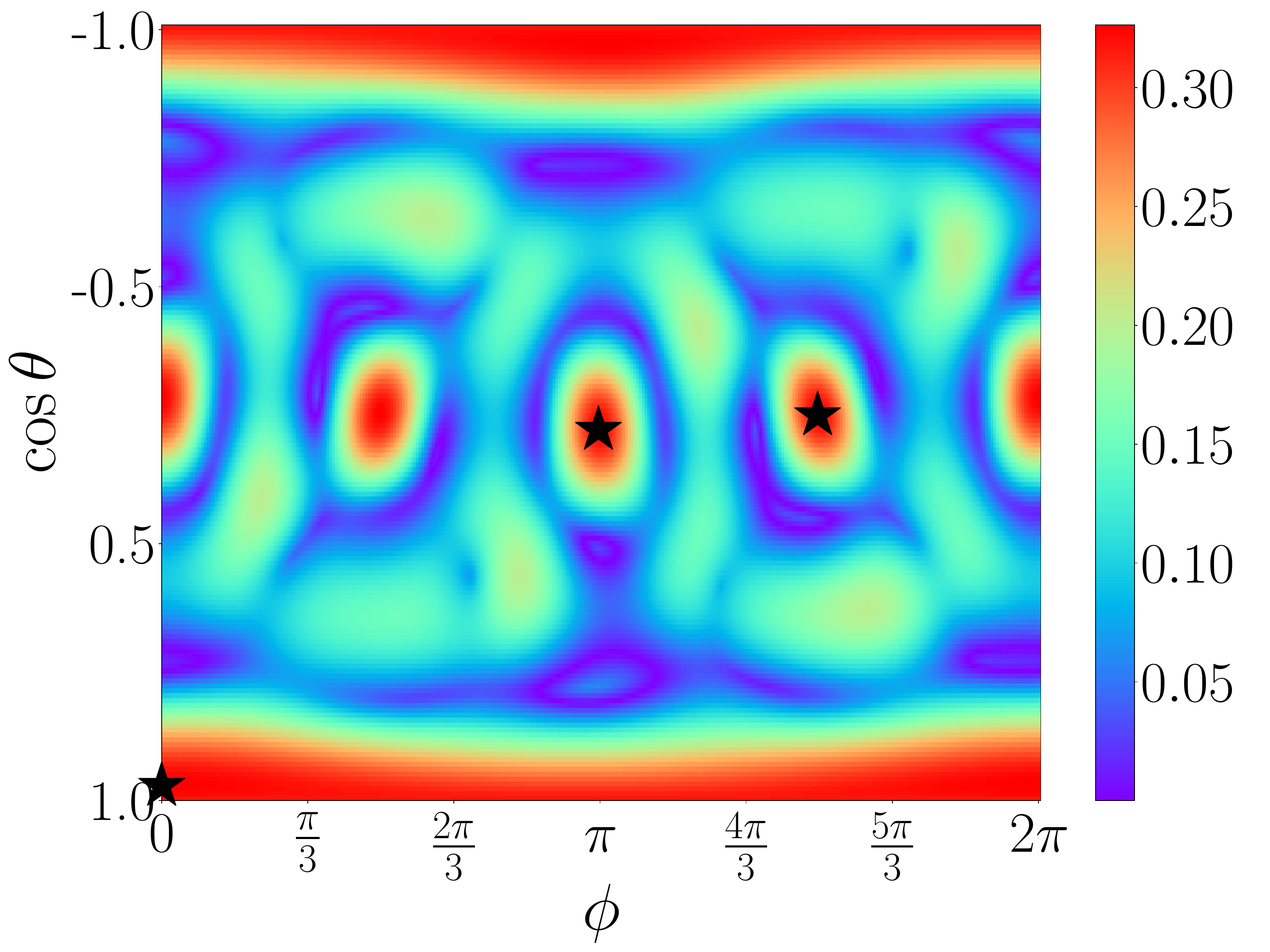}
    \caption{(A) Simulated sample of isotropic particles melted from an ideal FCC lattice. From this perspective, several possible domains can be seen. \newtext{Full 3D visualization can be generated from XYZ file included in Supplementary materials. } (B) A heat map representation of SymBOP signal vs. reference axis direction $\nu$  as given by Eq.~(\ref{eq:cubic_SymBOP_max_psi}), applied to this simulated sample. The red regions have the highest SymBOP signals, and they show a clear demonstration of cubic symmetry for this sample. Three orthogonal hot spots are chosen as the symmetry axes of the largest domain, labeled by black stars in the figure. }
    \label{fig:artificial}
\end{figure}

 First, we analyze the simulated sample of isotropic particles seen in Fig.~\ref{fig:artificial}A. Following this, we look at two experimental systems  composed of tetrahhedral DNA frames and isotropic DNA-grafted nanoparticles (NPs). In Ref.~\citenum{mushnoori2022controlling} we have analyzed simulated hybrid systems of patchy particles with interactions mediated by isotropic particles. The examples shown here are an extension of that work to include data without the orientational information of the constituent particles.


The simulated data were produced using the open-source molecular dynamics package LAMMPS. An ideal FCC lattice with 3D free boundaries was sampled in the canonical ensemble. The temperature was initially spiked and slowly cooled while the particles interacted through a Lennard-Jones potential. The final configuration consists of several different domains with varying orientations. The simulation is useful for two main reasons: it provides an example of using SymBOPs for a simulated sample of isotropic particles, and it demonstrates the ability of SymBOPs to detect many domains with different relative orientations.

The first step is to find the domain orientations relative to a global coordinate system, assuming that the underlying symmetry is cubic.  The largest ordered domain would correspond to a specific  optimal orientation that produces the highest SymBOP signal $\left|Q^*_l(\hat{\nu},\psi)\right|$, given by   Eq.~(\ref{cubic_SymBOP}). By  explicitly maximizing its value  with respect to axial rotation $\psi$, one  obtains the strength of the cubic signal parameterized only by orientation of the main axis $\hat \nu$:
\begin{equation}
\label{eq:cubic_SymBOP_max_psi}
Q^*_l(\hat \nu)\equiv \max\limits_{\psi}\left|Q^*_l\right|=|Q_l^4(\hat \nu)|\sqrt{2(1-\chi_l^2)} +\left|\chi_lQ_l^0(\hat \nu)\right|
\end{equation}

\begin{figure}[htb!]

       \centering
       \begin{flushleft}{A} \end{flushleft} 
    \includegraphics[width=0.6\linewidth]{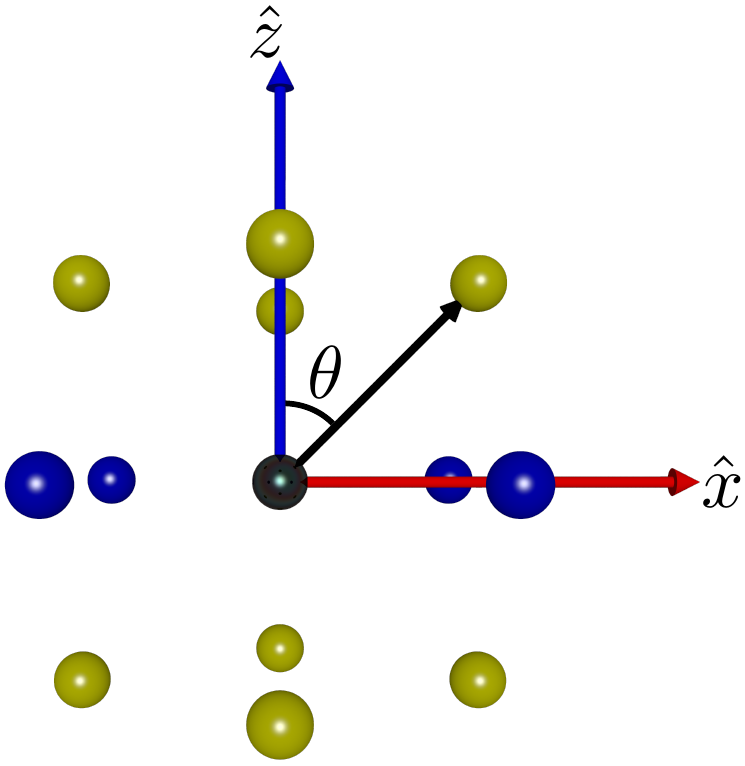}
    \includegraphics[width=\linewidth]{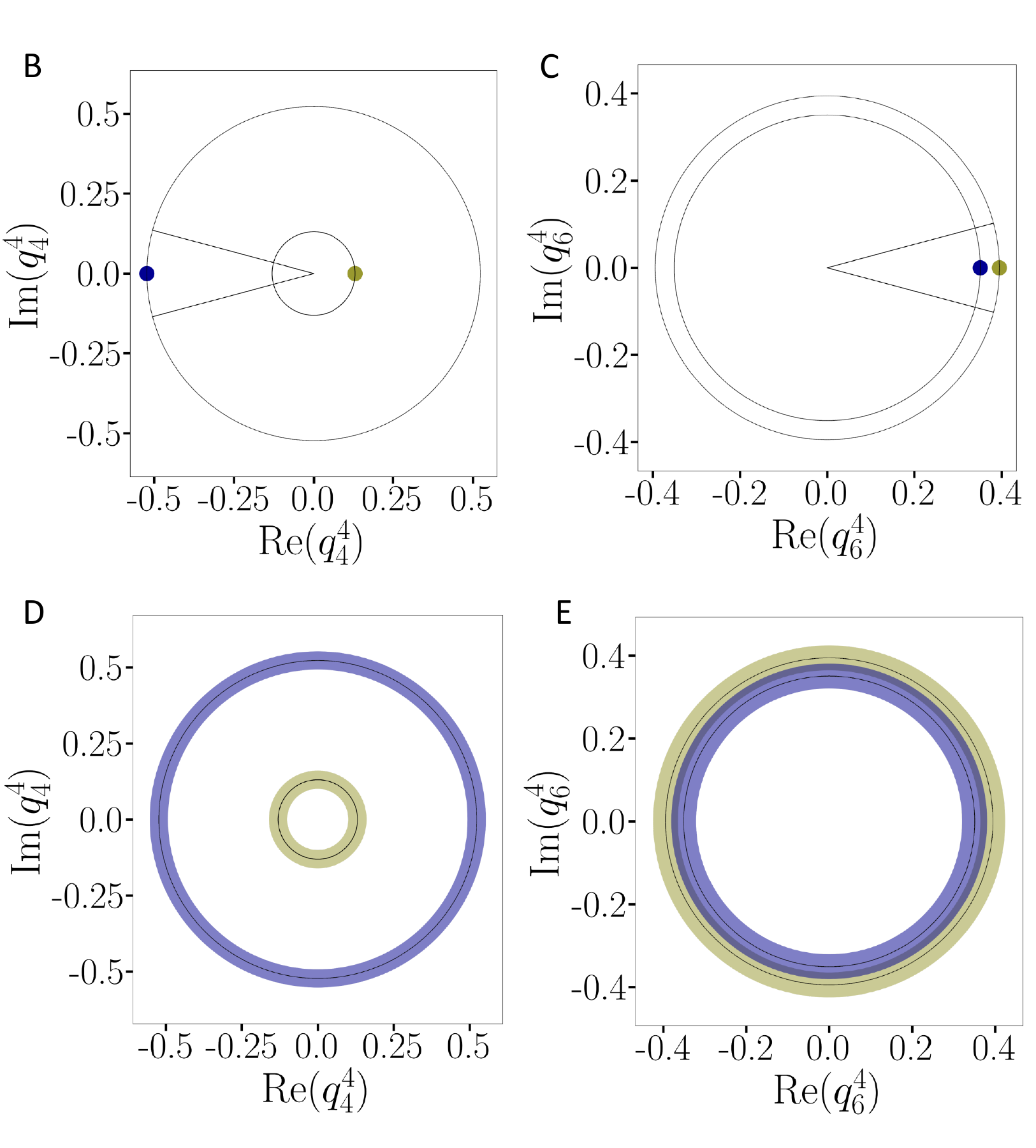}
    \caption{(A) The nearest neighbors of a particle in an ideal FCC crystal. There are two different types of bonds, up to rotation by $\phi=\pi/2$, colored blue and yellow, corresponding to $\theta=\pi/4$ and $\theta=\pi/2$. (B)-(C) The complex plane of the non-trivial component $q_{l}^4$ for $l=4,6$. Each bond in the sample will have a unique point in this plot. The points shown here represent the ideal bonds in an FCC neighborhood, and they are colored the same as their bond types in (A). (D)-(E) The remaining rotational degree of freedom spreads the ideal points into ideal rings.}
    \label{fig:ql4_ideal_FCC}
\end{figure}

Figure~\ref{fig:artificial}B represents this function for the simulated system, as a heat map in spherical coordinates. Specifically, for every direction of vector $\nu$, parameterized by two Euler angles, the color represents the value of SumBOP signal $Q^*_l(\hat \nu)$.     Each hot spot corresponds to the orientation of the symmetry axis of an ordered domain.  In the figure, the coordinate system is aligned with the axes of the largest domain that correspond to the strongest SymBOP signal. Naturally, there are six equivalent orientations, as expected for the cubic symmetry.

Once the orientations of the symmetry axes are determined, we are able to identify groups of bonds that are mutually aligned upon the symmetry transformation.  The  bond-level SymBOPs associated with both tetrahedral and cubic symmetries  may act as descriptors for these  classes of equivalent bonds:
\begin{align}
q^4_l({\hat b}_{ij})&=\left({\hat b}_{ij}|\hat{e}^4(\hat \nu_0,\psi_0)\right)_l \\
    |b_{ij}^*|_l=&\left|\left({\hat b}_{ij}|\hat{R}_c(\hat \nu_0,\psi_0)\right)_l\right|=\\
    \nonumber =&\sqrt{2(1-\chi_l^2)}\left|\Re (q_l^4({\hat b}_{ij}))\right|+|\chi_l q_l^0({\hat b}_{ij})|
\end{align}

As the tetragonal SymBOP $q_l^4$ is a complex number, it is generally more informative than the real-valued cubic bond descriptor $|b_{ij}^*|_l$. Thus we will use $q_l^4$ for bond classification and will discuss $|b_{ij}^*|_l$ later in the manuscript.  For a given crystal structure, each type of bond present will be represented by a point in the complex plane of $q^4_l$. For instance, Fig.~\ref{fig:ql4_ideal_FCC}A shows the nearest neighbors in an ideal FCC crystal. For FCC there are only two types of bonds: those that lie at $\theta=\pi/4$ and $\theta=\pi/2$. Figure~\ref{fig:ql4_ideal_FCC}B-C show $|q_l^4)$ plots with the two types of FCC bonds labeled with their respective color. To find the final angle necessary to fix the reference vector we could plot $|q_l^4(\hat b_ij)$ for each bond. If the first two Euler angles have been chosen correctly, we expect to see clustering of the bonds, showing that they have similar orientations. However, because there is a single rotational degree of freedom remaining, these ideal points spread into ideal rings, shown in Fig.~\ref{fig:ql4_ideal_FCC}D-E, and we can only be certain about the radius at which the ideal bonds should lie. The blue and yellow rings identify the regions of near-ideal FCC bonding. These plots allow us to fully capture the bonds present in the sample. Because the sample is not ideal, we expect the bonds that belong to an ordered structure to spread into clusters of points that fall near the radii of the ideal rings. However, we cannot be certain at what angular position around the rings the ideal bonds will lie. The clustering can then be captured with a histogram of the points that fall in the colored rings of the plots. The peak of the histogram or kernel density estimate will locate the densest region of bonds around the ideal ring.

\begin{figure}[htb!]
    \centering
\begin{flushleft}{A} \end{flushleft} 
\includegraphics[width=0.7\linewidth]
    {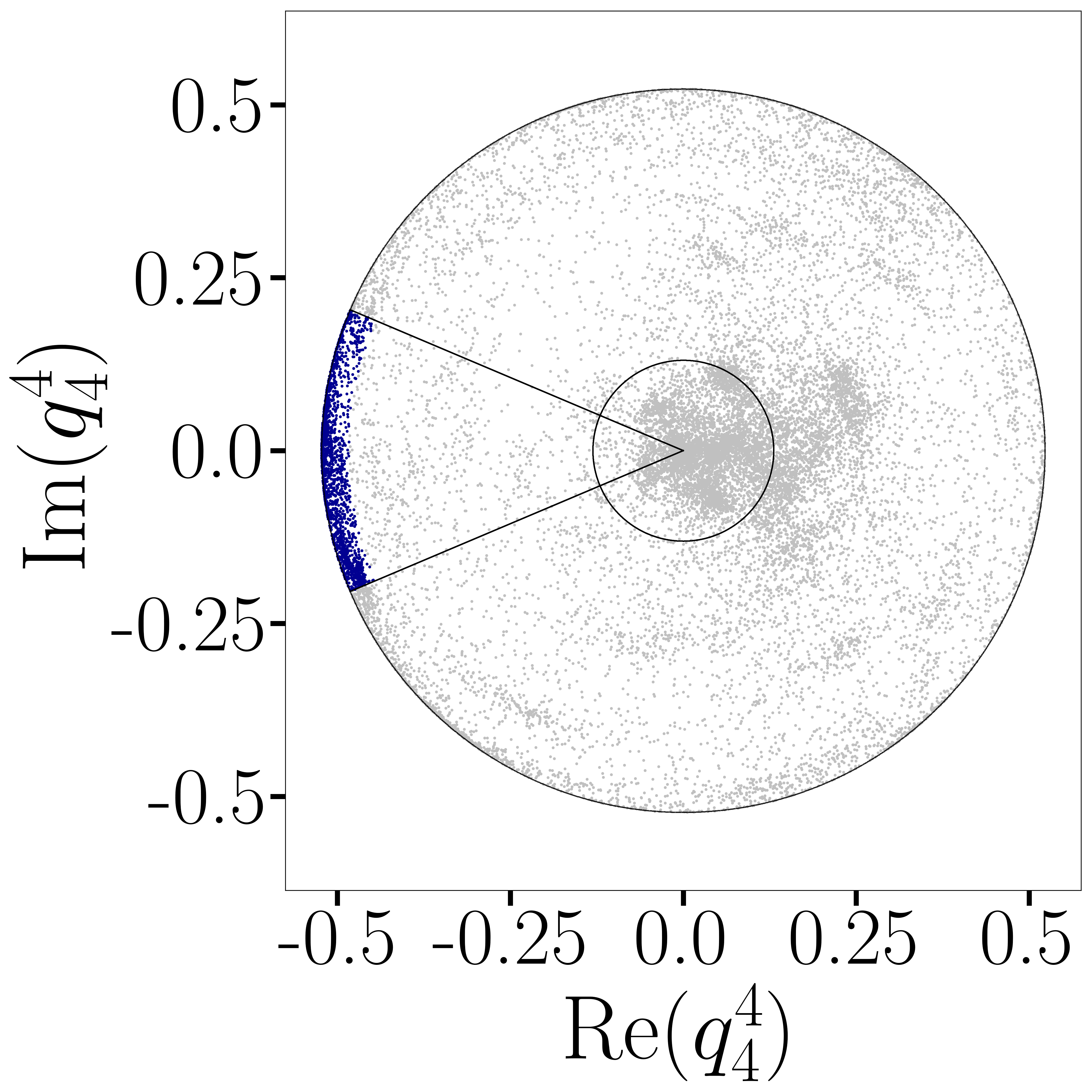}
    \begin{flushleft}{B} \end{flushleft} 
\includegraphics[width=0.7\linewidth]
    {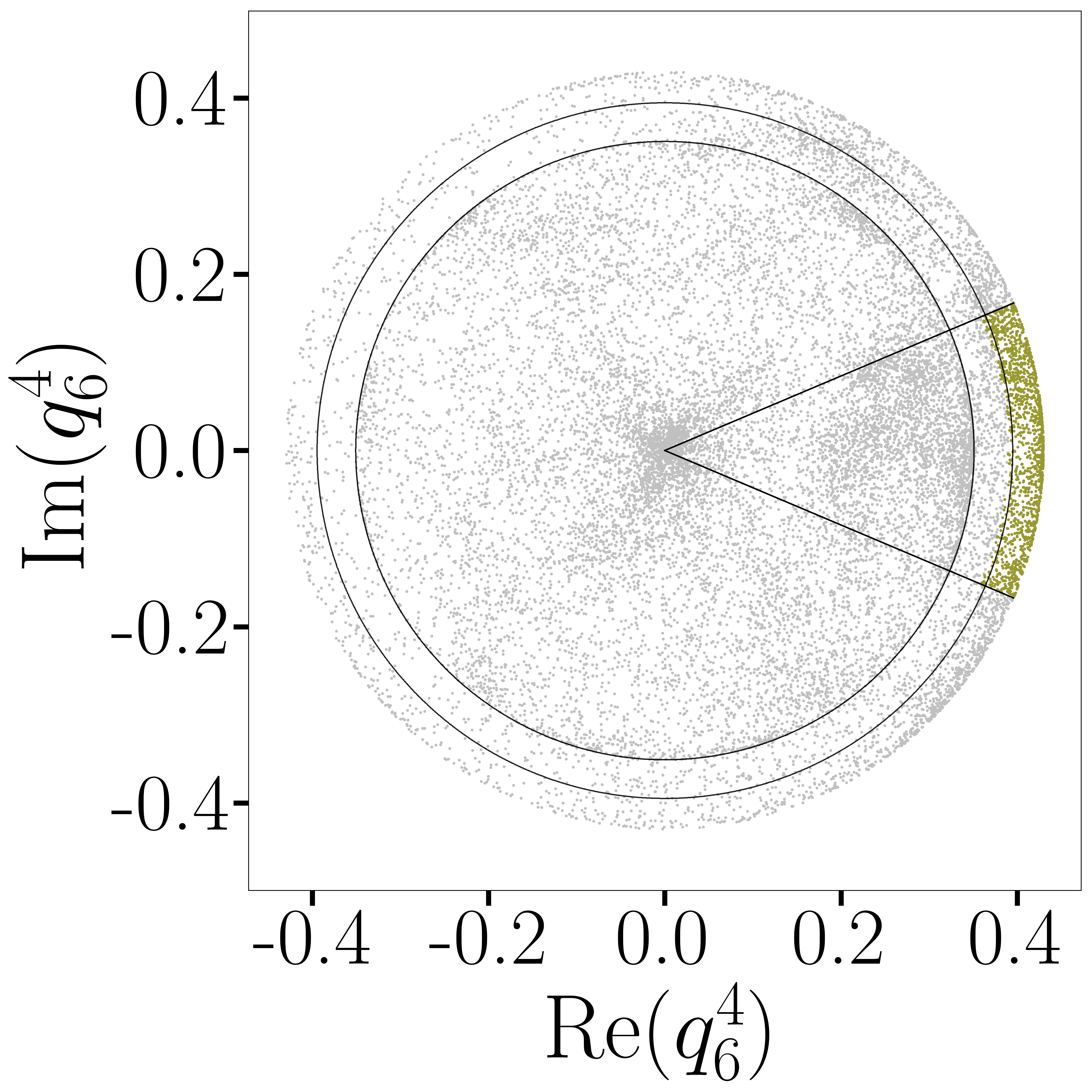}

    \caption{(A)-(B) Complex $q_{l}^4$ plots for the simulated sample. Here we have included the data collected for each type of bond, shown in blue and yellow, in both plots. This demonstrates that although all of the data can be collected in one of the plots, it is more convenient and accurate to use both plots for $l=4$ and $l=6$.}
        \label{fig:ql4_plots_artificial}
   \end{figure} 

\begin{figure}[htb!]
    \centering
\includegraphics[width=0.9\linewidth]
    {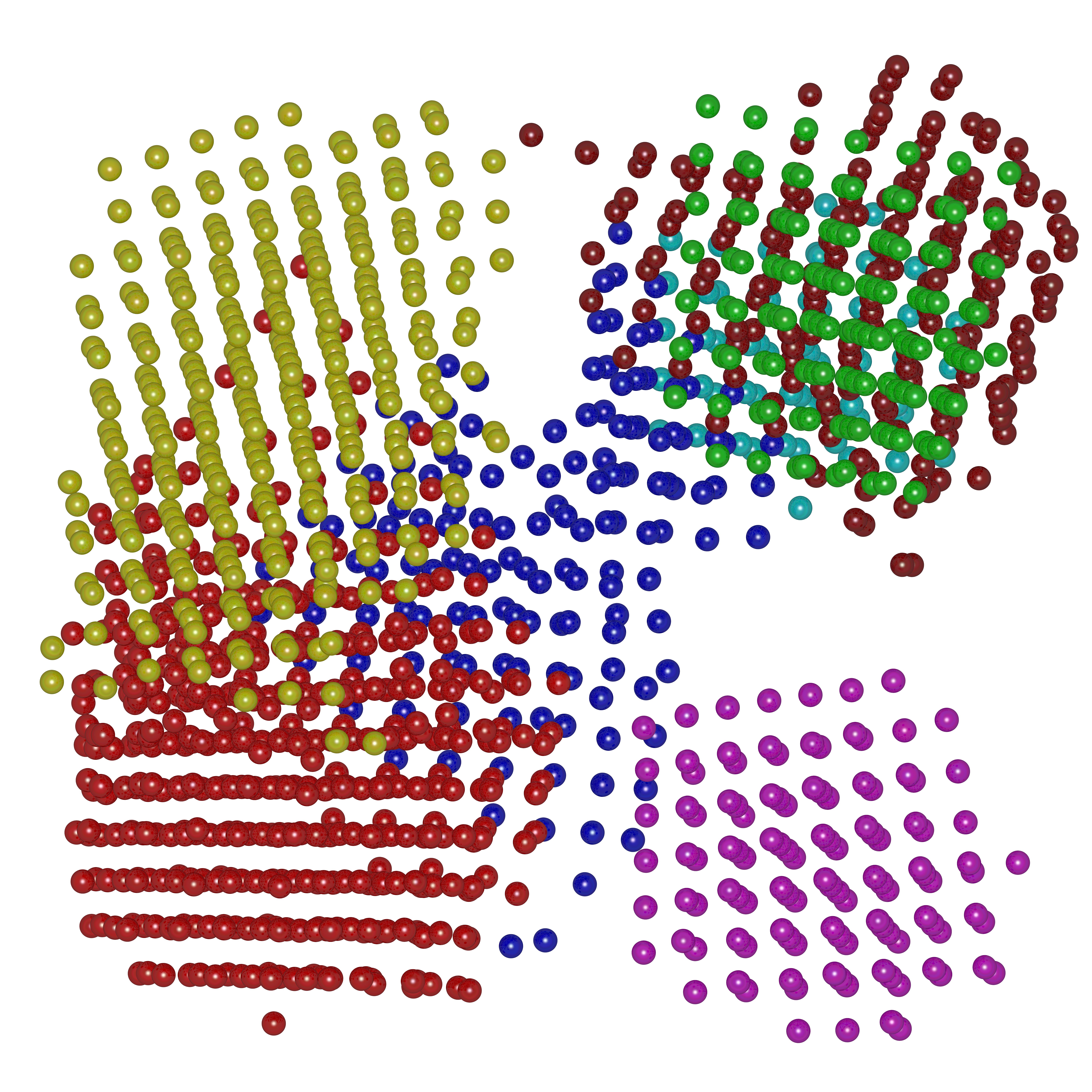}
\caption{The ordered domains were found inside of the simulated sample\newtext{, shown here as a 2D projection to emphasize their differing orientations}. Each color represents a spatially distinct domain, but several domains can be found for one orientation of the symmetry axes. \newtext{Full 3D visualization can be generated from XYZ file included in Supplementary materials. }}
    \label{fig:simulation}
\end{figure}

Alternatively, one can remove this last rotational degree of freedom by choosing not one, but three hot spots from Fig.~\ref{fig:artificial}B. In the case of cubic symmetry, these hot spots should represent orthonormal vectors. If a single hot spot aligns the $z$-axes of the reference vector and the global coordinates of the desired domains, three hot spots will completely align the coordinate systems. The black stars in Fig.~\ref{fig:artificial}B label the chosen hot spots for the sample. If the coordinates match, then the exact positions of the ideal bonds are known. For FCC, any near-ideal bonds should fall in the neighborhood of the points in Fig.~\ref{fig:ql4_ideal_FCC}B-C. This route only requires looking for bonds that fall in a neighborhood of the ideal points.

In Fig.~\ref{fig:ql4_plots_artificial}, the chosen bonds are shown in colors matching their bond type. These chosen bonds may not all be spatially localized even though they have similar orientations, which leads to the possibility of multiple independent domains with very similar relative orientations. After collecting the bonds we use bond percolation to identify the particles that belong to the same  domain.

The bond percolation procedure may in principle find several independent domains with the same orientation. The process can then be iterated to find all domains in the sample. To do this,  the bonds that belong to the identified domains are eliminated, and the procedure is repeated again. New optimal orientations of the reference vector can be found based on  the strongest SymBOP signal remaining in the sample. Figure~\ref{fig:simulation} shows the domains found for this simulated sample. In this sample, seven domains were identified spanning three unique orientations.

\subsection{\label{subsec:Experimental Data}Domains and defects in experimental self-assembled structures}

As a further demonstration of the SymBOP method, we apply the procedure described above to experimentally self-assembled NP lattices. Both samples that are discussed below were assembled using DNA-origami tetrahedral frames combined with DNA-grafted gold NPs \cite{diamond_oleg,michelson2022three}. The self-assembled superlattices were encased in radiation-hard silica and imaged using Scanning Hard X-ray Microscopy with a nanobeam. The resultant 3D tomography of the gold positions was resolved into spatial coordinates of each NP in the structure. The experimental methodology,  previously introduced in \citenum{michelson2022three}, is overviewed in Appendix A. In the first example, DNA frames bind to NPs via vertices that contain single-stranded(ss) DNA complementary to  ssDNA grafted to NP surfaces.  In the second case, the tetrahedral DNA frames bind directly to each other via their vertices and have additional ssDNA pointing inward. That  allows encapsulating gold NPs grafted with complementary ssDNA,  inside the frames. 



\begin{figure}[htb!]
    \centering
    \begin{flushleft}{A} \end{flushleft} 
    \includegraphics[width=0.6\linewidth]{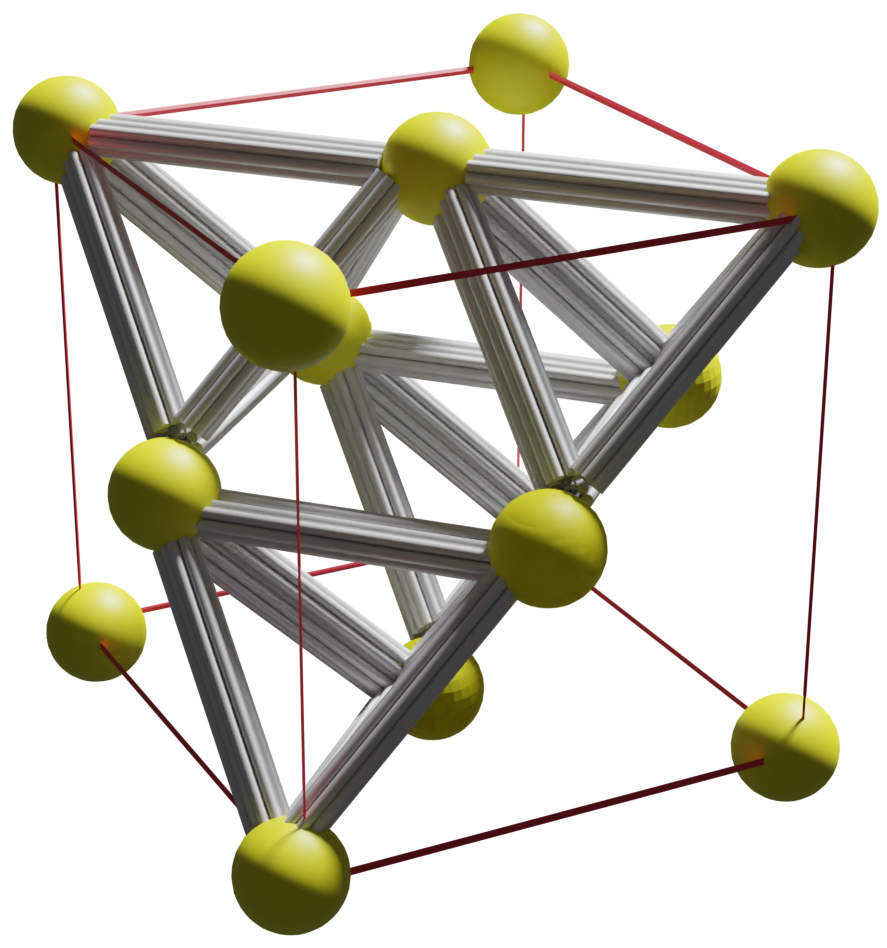}
     \begin{flushleft}{B} \end{flushleft} 
    \includegraphics[width=0.8\linewidth]{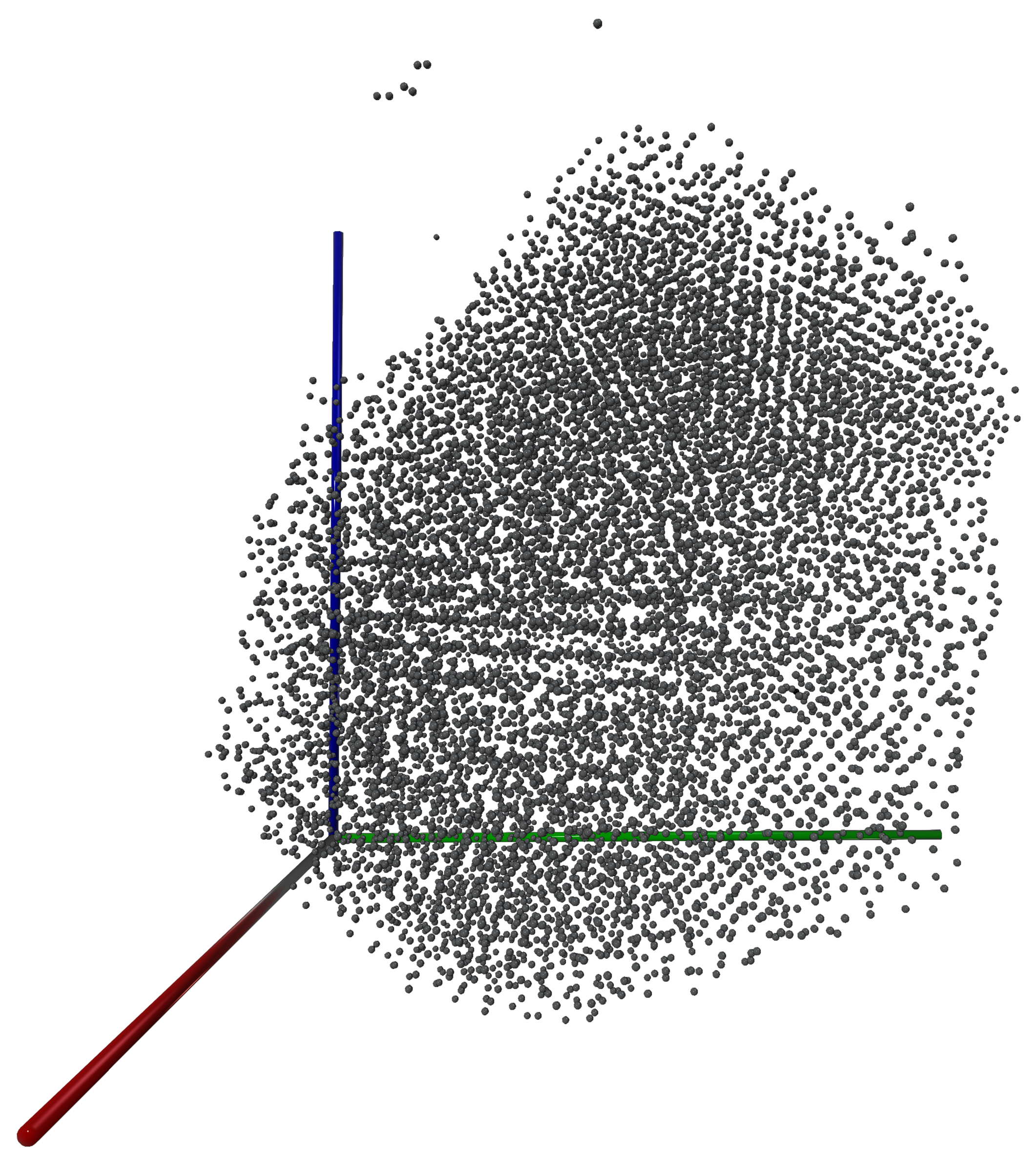}
    \caption[]{ (A) An illustration of the NPs binding to tetrahedral DNA frames to self-assemble in one cell of an ideal FCC lattice. The gray bundles depict the DNA frames, and the yellow spheres represent the ssDNA-grafted gold NPs, which are seen to be bound to the complementary ssDNA at the vertices of the frames. \newtext{(B) A 2D projection of an} experimental sample of tetrahedral frames with gold NPs bound to their vertices. Because we lack the local orientational information of each frame, they are visualized as isotropic particles. \newtext{ Full 3D visualization can be generated from XYZ file included in Supplementary materials.}}
    \label{fig:exp_HXN_sample}
\end{figure}

\begin{figure}[htb!]
    \centering
\begin{flushleft}{A} \end{flushleft} 

    \includegraphics[width=\linewidth]{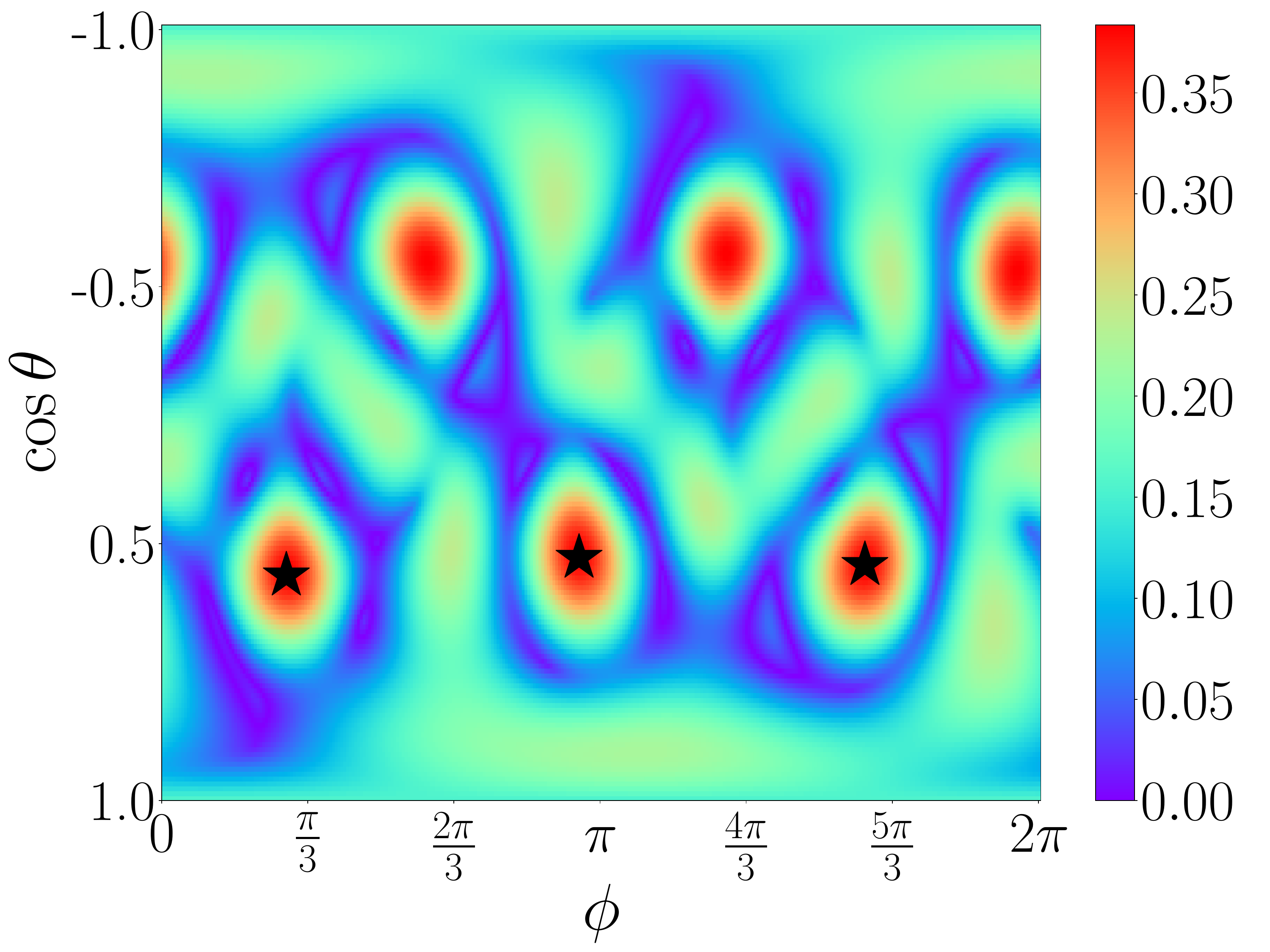}
\begin{flushleft}{B} \end{flushleft} 
    
    \includegraphics[width=\linewidth]{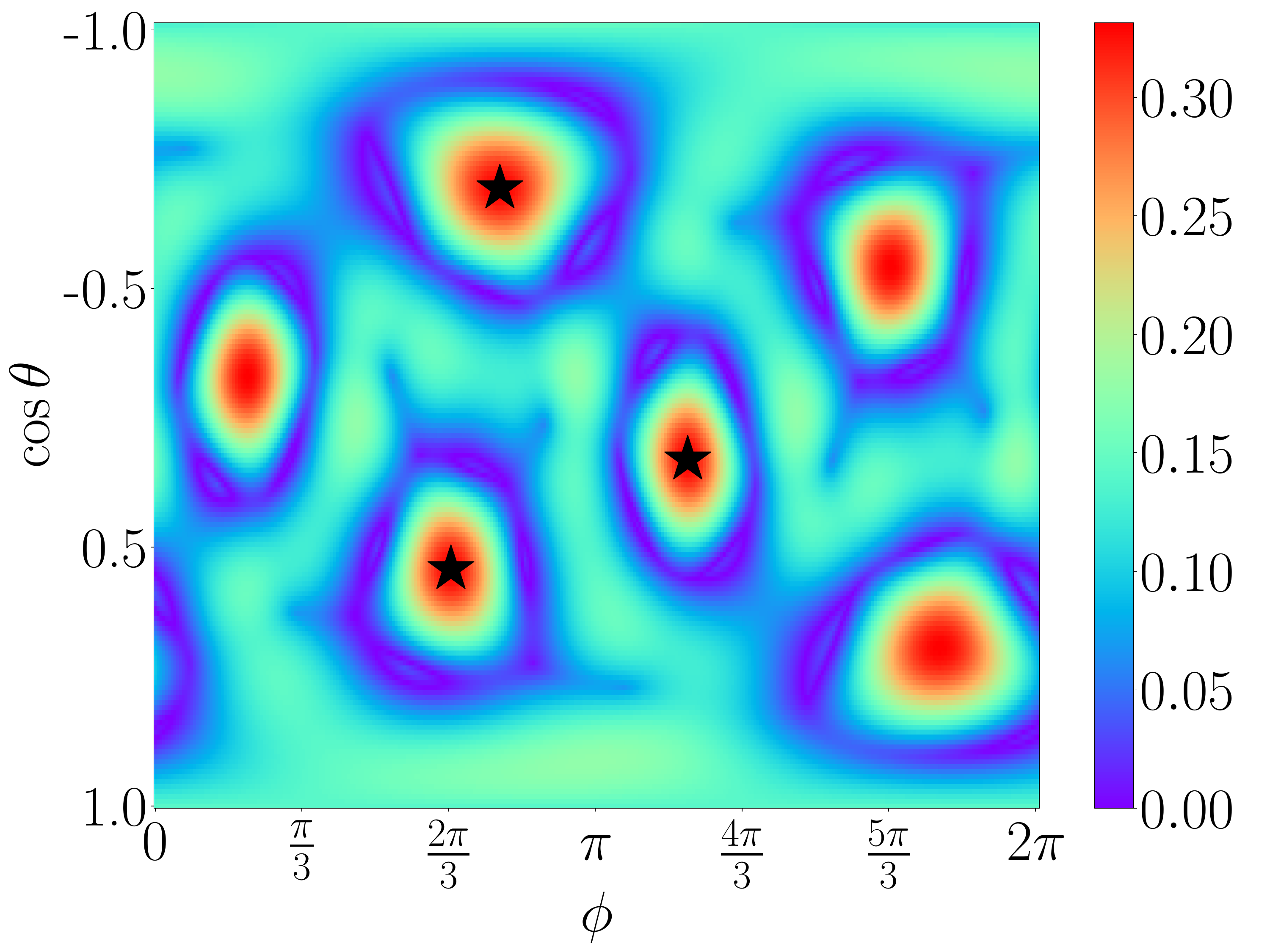}
    \caption[]{(A) A heat map representation of Eq.~(\ref{eq:cubic_SymBOP_max_psi}), as applied to the experimental sample with NPs bound to the vertices of the DNA frames. The red regions have the highest SymBOP signals, and three orthogonal points are chosen as the symmetry axes (black stars) of the first domain. (B) Heat map showing the hot spots after removing the bonds that are part of  the first discovered domain. Choosing three orthogonal axes, marked with black stars, the orientation of the second domain with cubic symmetry can be found.}.
    \label{fig:exp_10_23_HXN_heatmaps}
\end{figure}

\subsubsection{Identification of coherent domains \label{sec:two domain experimental data}}

\begin{figure}[htb!]
    \centering
    \includegraphics[width=1\linewidth]{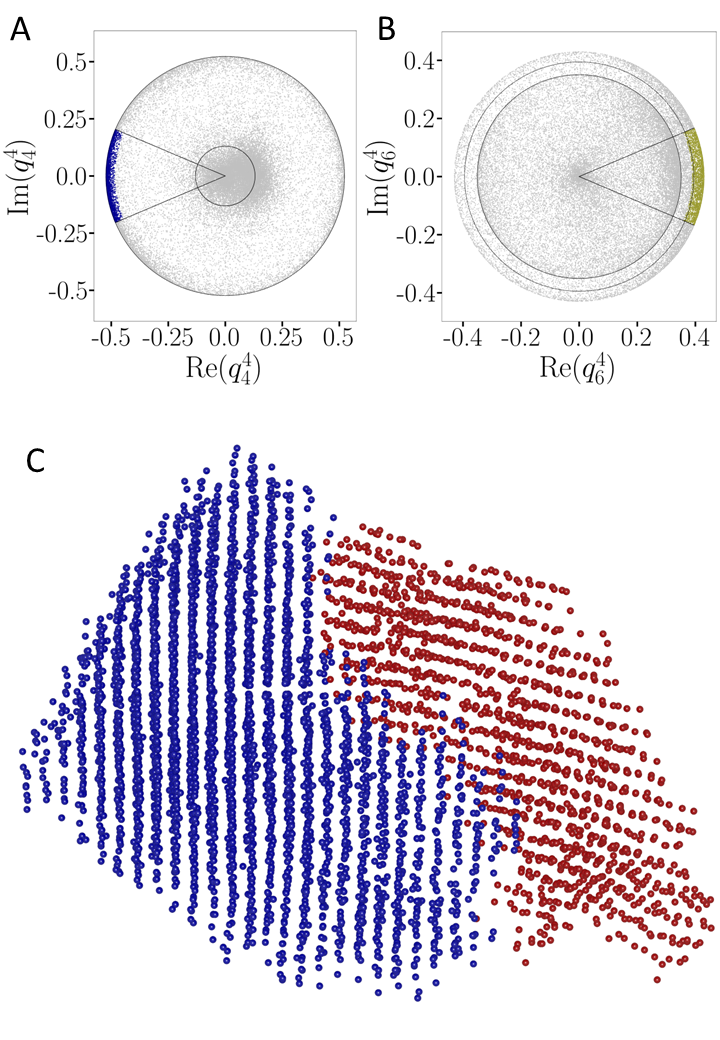}
    \caption{Analysis of an experimentally self-assembled FCC lattice gold NPs bound to the vertices of tetrahedral DNA frames. (A-B) $q_l^4$ complex planes for $l=4,6$. Bonds in a neighborhood of the ideal points, shown in Fig.~\ref{fig:ql4_ideal_FCC}B, are chosen. (C) \newtext{An orthographic view of the 3D domains} found from percolating the bonds chosen in (A-B). Here there are two main domains. From the image it is clear that their orientations are not the same. \newtext{ Full 3D visualization can be generated from XYZ file included in Supplementary materials.}}
    \label{fig:exp_HXN_ql4_plots}
\end{figure}

\begin{figure}[htb!]
    \centering
\begin{flushleft}{A} \end{flushleft} 
    \includegraphics[width=0.75\linewidth]
    {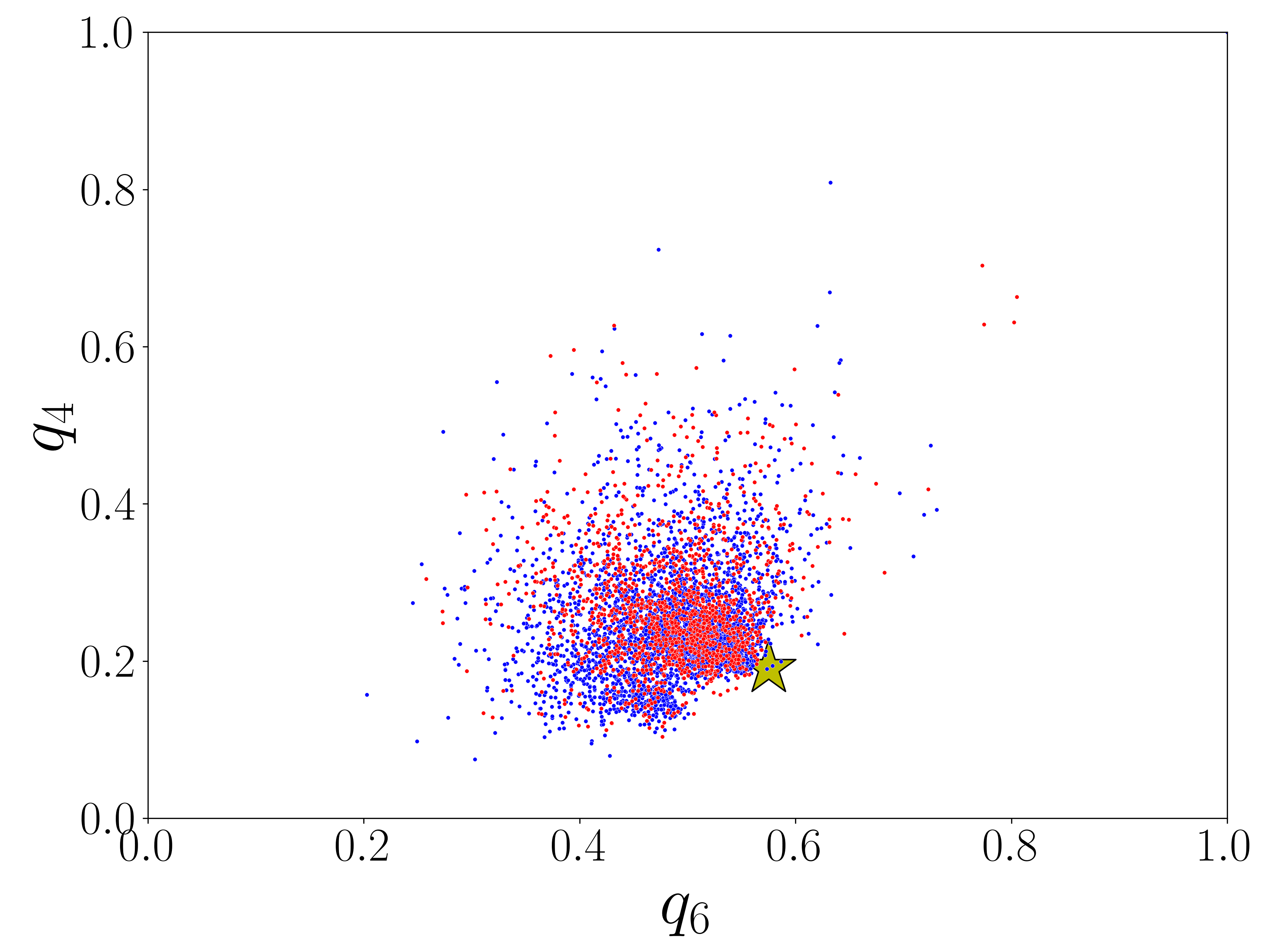}
    \begin{flushleft}{B} \end{flushleft} 
    \includegraphics[width=0.75\linewidth]
    {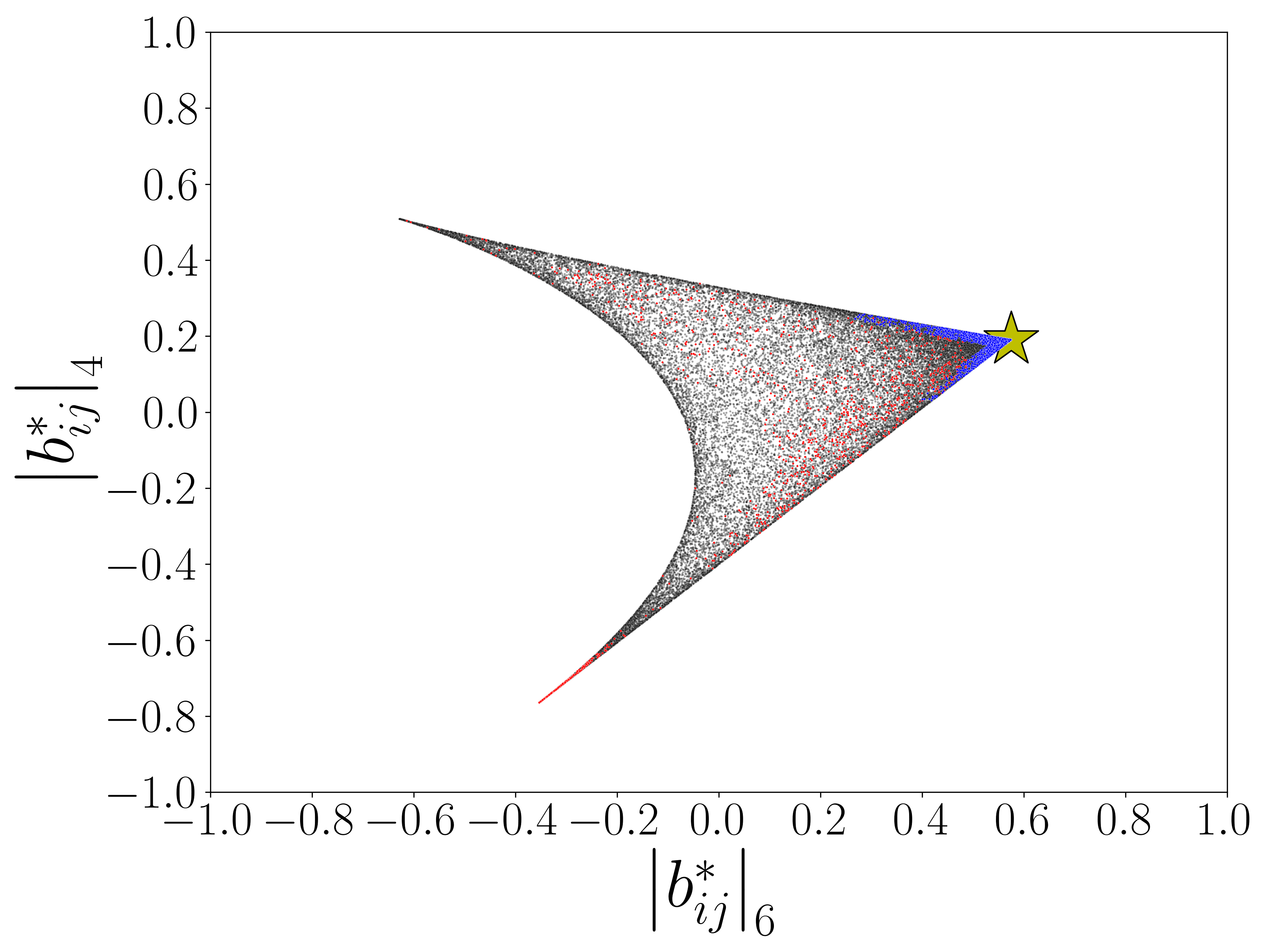}
    \begin{flushleft}{C} \end{flushleft} 
    \includegraphics[width=0.75\linewidth]
    {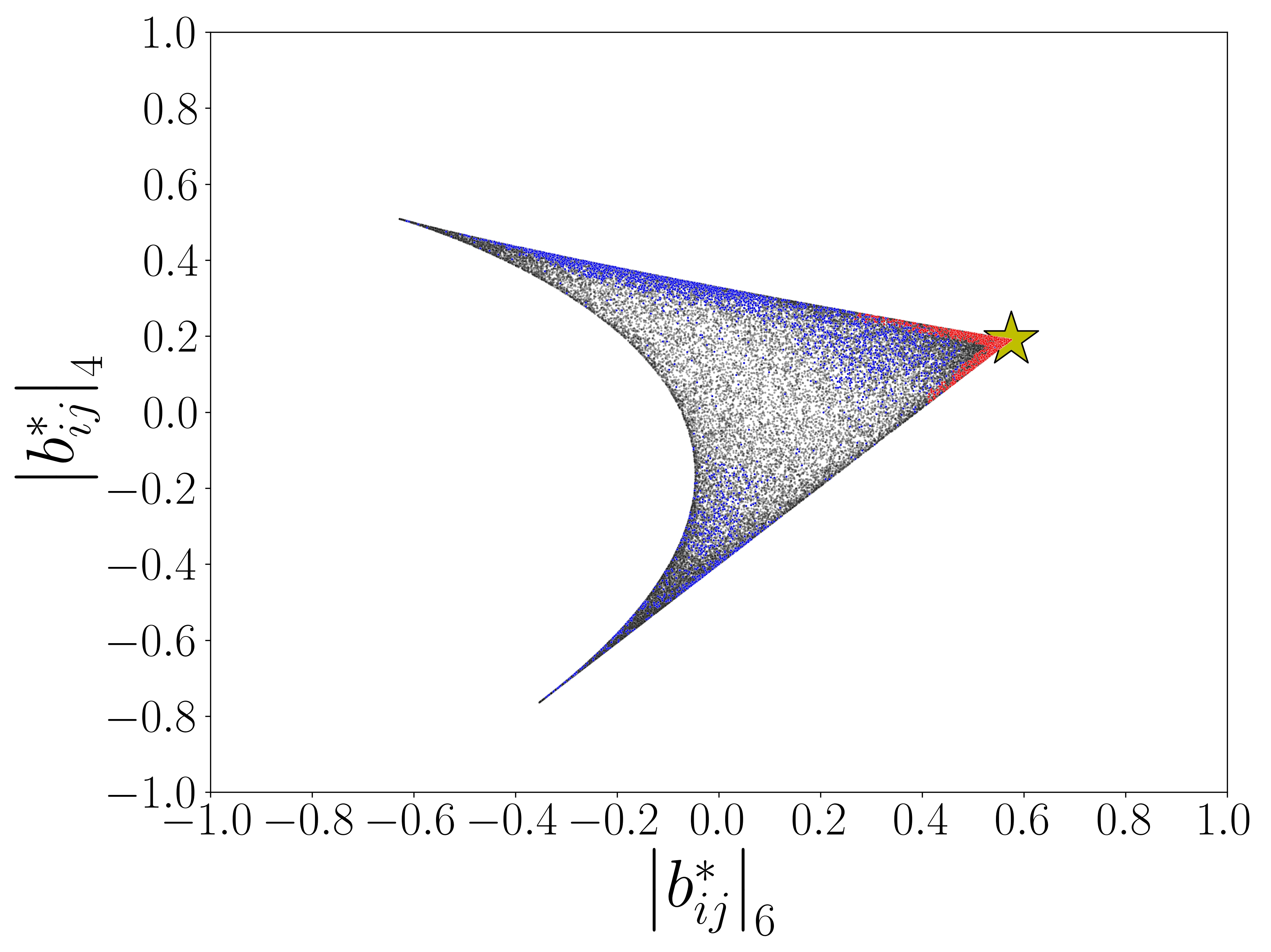}
    \caption[]{Scatter plots comparing the classical scalar BOPs and SymBOP analysis. The star on all panels indicates the location of the ideal FCC crystal in $(q_6,q_4)$ coordinates. (A) Particle-level data of scalar BOPs $(q_4,q_6)$ which are commonly used for characterization of local ordering.  As one can see, the particles that belong to the ``red" and ``blue" domains (represented by data points of the corresponding colors) are not very well clustered and  could not be distinguished with this method. (B)-(C) Bond-level SymBOPs given by Eq.~(\ref{eq:bond_level}), for two reference orientations that correspond to the principle directions of the ``blue" and ``red" domains, respectively. In each case, the bonds that belong to the respective domains are clearly identifiable as clusters of data points next to the ideal FCC location (the star).} \label{fig:exp_10_23_HXN_bond_symbop_plots}
\end{figure}

Figure~\ref{fig:exp_HXN_sample}A shows the idealized image in which the DNA-grafted gold NPs bind to vertices of DNA nancoges to form an ordered FCC crystal. Fig.~\ref{fig:exp_HXN_sample}B  shows the actual experimental sample.
We begin our analysis by finding the appropriate reference vector. Guided by our prior knowledge about the underlying structure, we are looking for domains with cubic symmetry. In a more general case, one would need to try a number of different SymBOPs corresponding to different symmetry groups, and identify those one that yield statistically significant signals.   The  reference vector for cubic SymBOP takes the form given by  Eq.~\ref{eq:ref_vector_cubic_symmetry}, but one must find the orientation of the coordinate system that matches domains that are present in the sample. Using the SymBOP projection we search for the highest signal. Figure~\ref{fig:exp_HXN_ql4_plots}A shows the heat map of this signal as a function of two of the Euler angles, $\phi$ and $\theta$, as given by Eq.~(\ref{eq:cubic_SymBOP_max_psi}). The hot spots define the preferred orientation for the coordinate system. After that, we are using the $q_l^4$ analysis in combination with the bond percolation procedure to identify the domain associated with this orientation. Following this, the bonds that belong to the first domain are eliminated from the analysis, and the entire procedure is repeated again.  Fig.~\ref{fig:exp_10_23_HXN_heatmaps}B is showing the SymBOP signal for the second iteration.   On each iteration, the scatter plots for bond-level  $q_l^4$ shown  in Fig.~\ref{fig:exp_HXN_ql4_plots}A-B are constructed to identify the clusters and perform the bond percolation procedure. 

The domains found by this method are colored ``blue" and ``red" in Fig.~\ref{fig:exp_HXN_ql4_plots}C. Iteration of the method allowed us to identify each individual domain, including the boundary particles that do not have a complete set of the nearest neighbors from the same domain.

\begin{figure}[htb!]
    \centering
    \begin{flushleft}{A} \end{flushleft} 
    \includegraphics[width=0.8\linewidth]{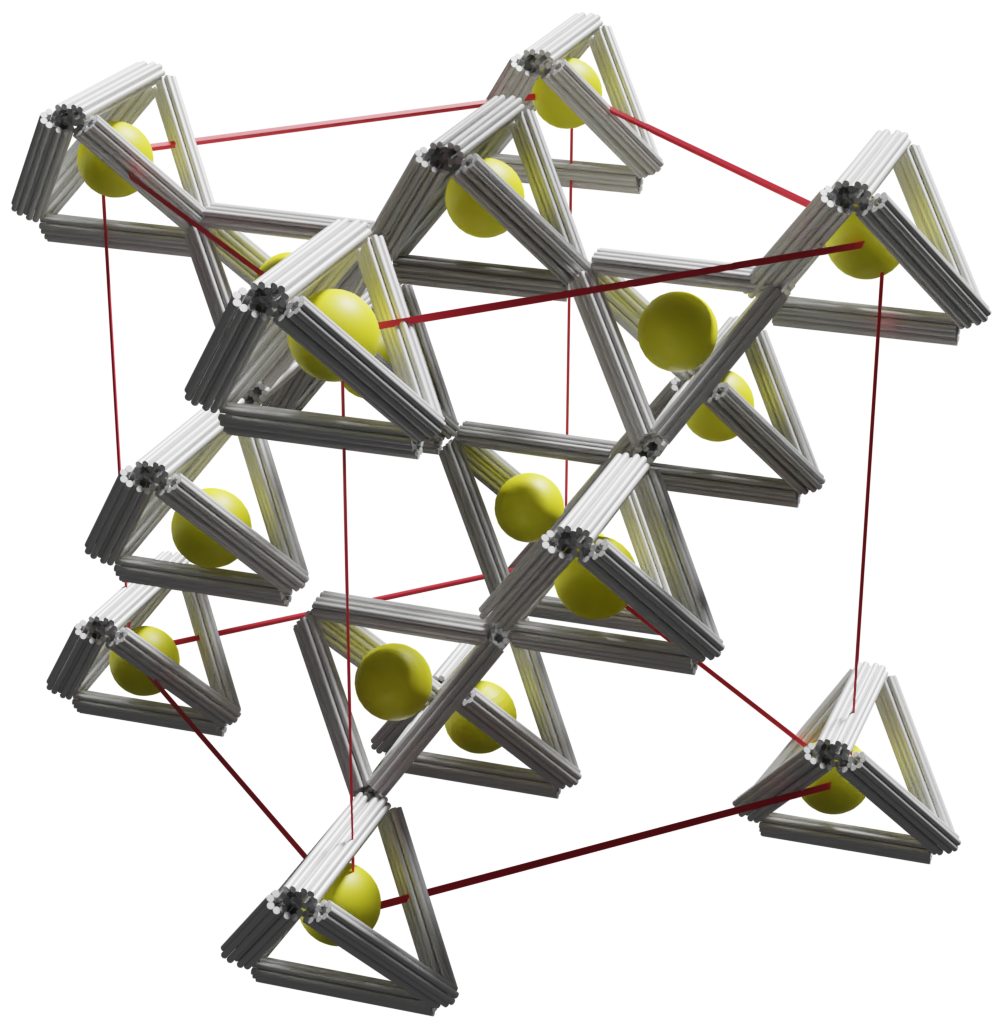}
    \begin{flushleft}{B} \end{flushleft} 
    \includegraphics[width=1\linewidth]{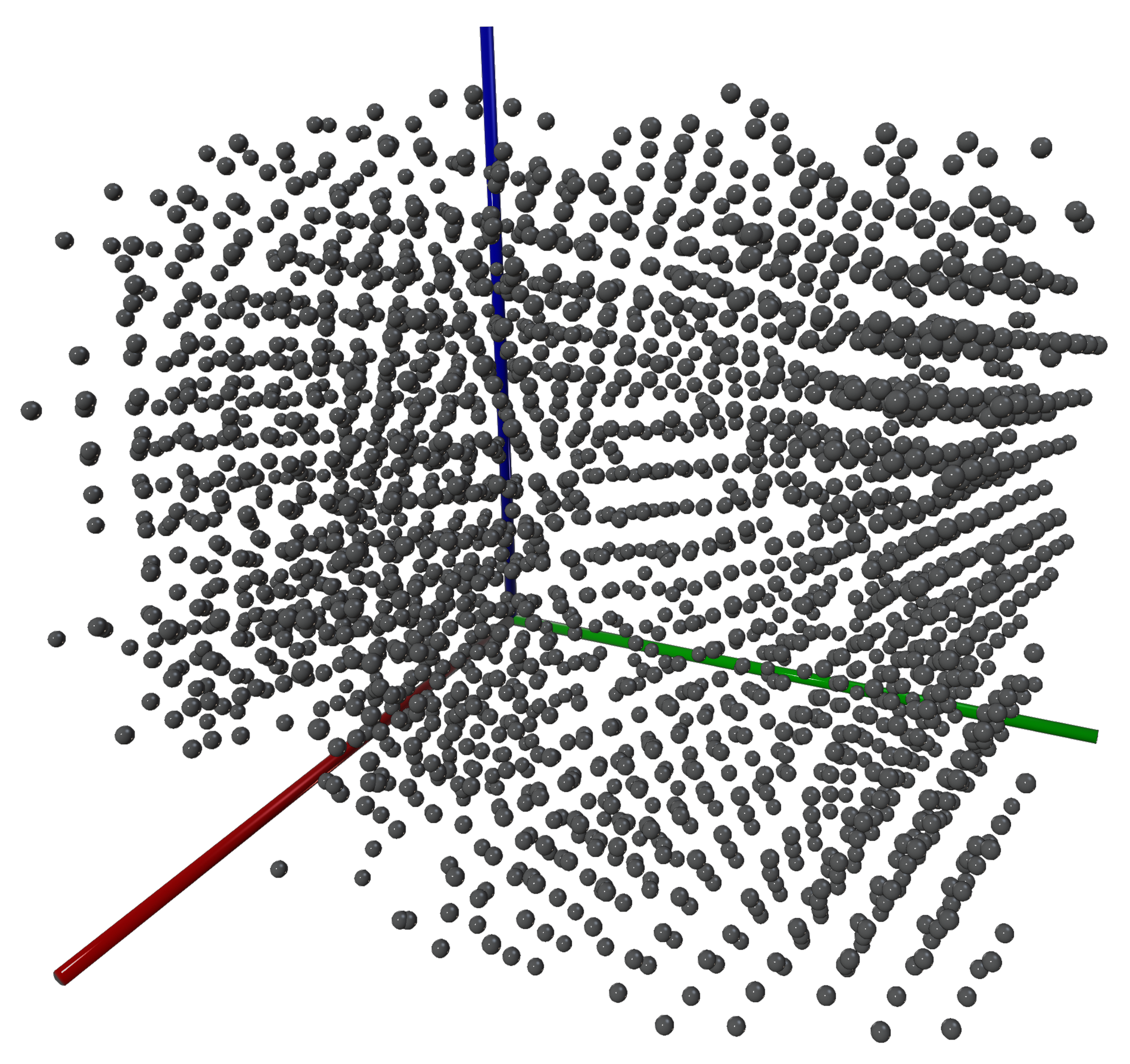}
    \caption[]{(A) An illustration of the DNA frames self-assembled in an ideal cubic diamond lattice. The gray bundles depict the DNA frames, and the yellow spheres the gold NPs encapsulated inside each frame. (B) Macroscopic view of the experimental sample made by self-assembling the NPs shown in (A). Because we lack the local orientational information of each frame, they are visualized as isotropic particles.\newtext{ Full 3D visualization can be generated from XYZ file included in Supplementary materials.}}
    \label{fig:exp_coh_bounds_sample}
\end{figure}

Figure~\ref{fig:exp_HXN_ql4_plots}C  makes it plain to see the sample has two main domains that come into contact along a large domain boundary. The SymBOP method allows us to find and study this domain boundary. 

The projection operator inherent in the SymBOP method has the advantage of providing information at the level of a single bond. In addition, the reference vector specifies the point symmetry group that we identify in the sample. These characteristics allow us to discover domains and ordered regions that standard BOPs cannot. 

Figure~\ref{fig:exp_10_23_HXN_bond_symbop_plots}A shows a $q_4$-$q_6$ plot, using traditional BOPs, of the domains found in the first experimental sample and displayed in Fig.~\ref{fig:exp_HXN_ql4_plots}C. The yellow star marks the point on the plot where particles with ideal FCC neighborhoods are expected to be found. Every point is a particle found in one of the ordered domains, colored accordingly. The points do not all fall where expected for particles in an FCC-like lattice. It's clear from the plot how challenging it would be to use the standard BOPs to discover each  domain. Furthermore, it would be impossible to separate them.
In contrast, Figs.~\ref{fig:exp_10_23_HXN_bond_symbop_plots}B-C show scatter plots of the bond-level  SymBOPs $\left|b^*_{ij}\right|_l$, with the data point colors matching the corresponding  domains in Fig.~\ref{fig:exp_HXN_ql4_plots}C. In Fig.~\ref{fig:exp_10_23_HXN_bond_symbop_plots}B the symmetry axes of the large blue domain are aligned with the global coordinate system, by identifying the highest SymBOP signal in Fig.~\ref{fig:exp_10_23_HXN_heatmaps}A. The bonds in the blue domain fall exactly in the neighborhood of the standard BOP  values for ideal FCC $(0.57,0.19)$, while the red domain is not well resolved in this orientation. Figure~\ref{fig:exp_10_23_HXN_bond_symbop_plots}C, however, shows a similar scatter plot  after the coordinate system is realigned with the principle axes of the second domain (that is, by  maximizing the SymBOP signal without the contribution of the ``blue" domain). The SymBOP for each bond in the ``red" domain is now clustered near the ideal FCC values, while the ``blue" bonds are no longer clustered.

\subsubsection{Identification of defects\label{sec:coh boundary experimental data}}

\begin{figure}[htb!]
    \centering
    \begin{flushleft}{A} \end{flushleft} 
       \includegraphics[width=1\linewidth]
    {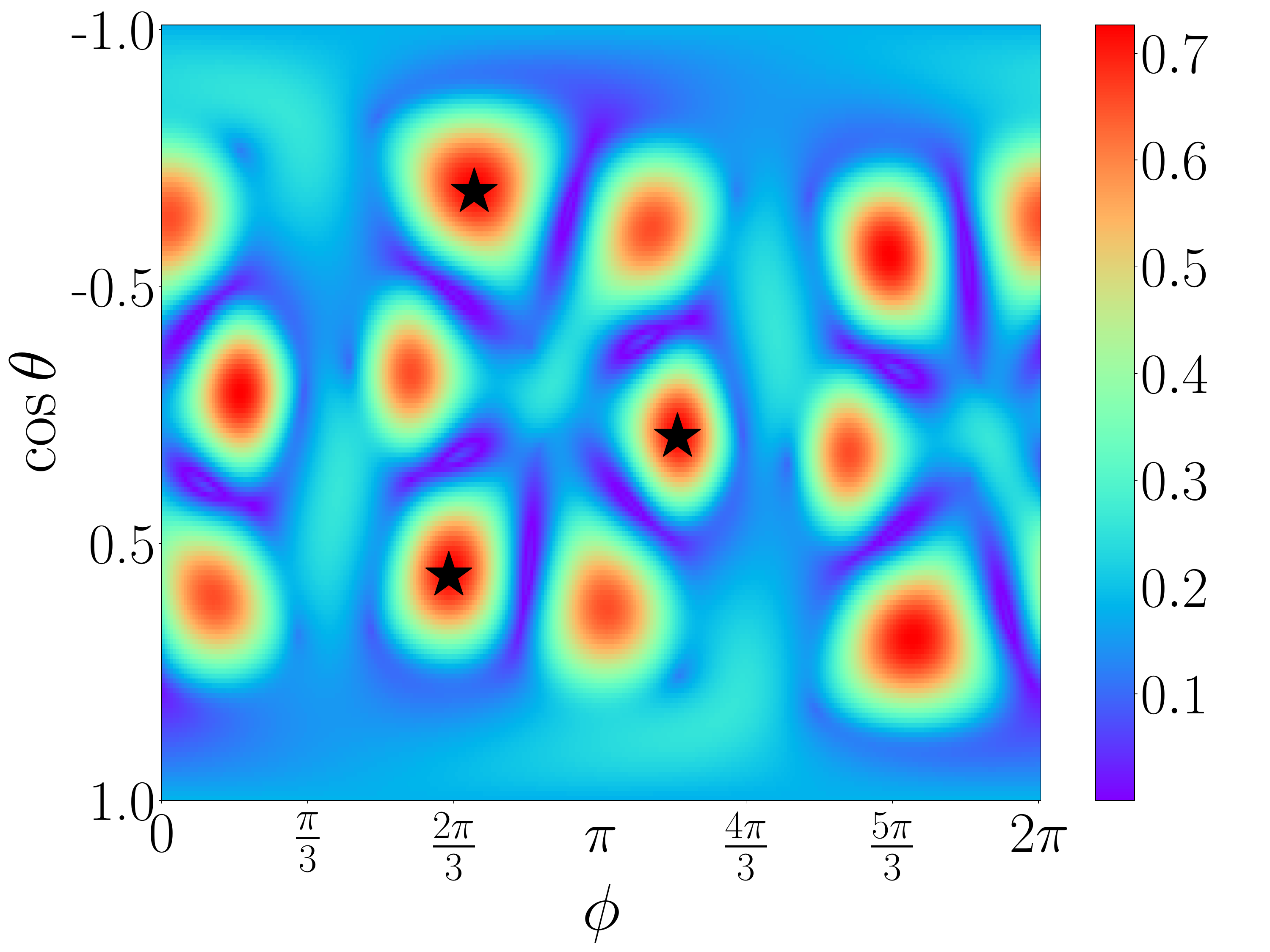}
      \begin{flushleft}{B} \end{flushleft} 
       \includegraphics[width=1\linewidth]
    {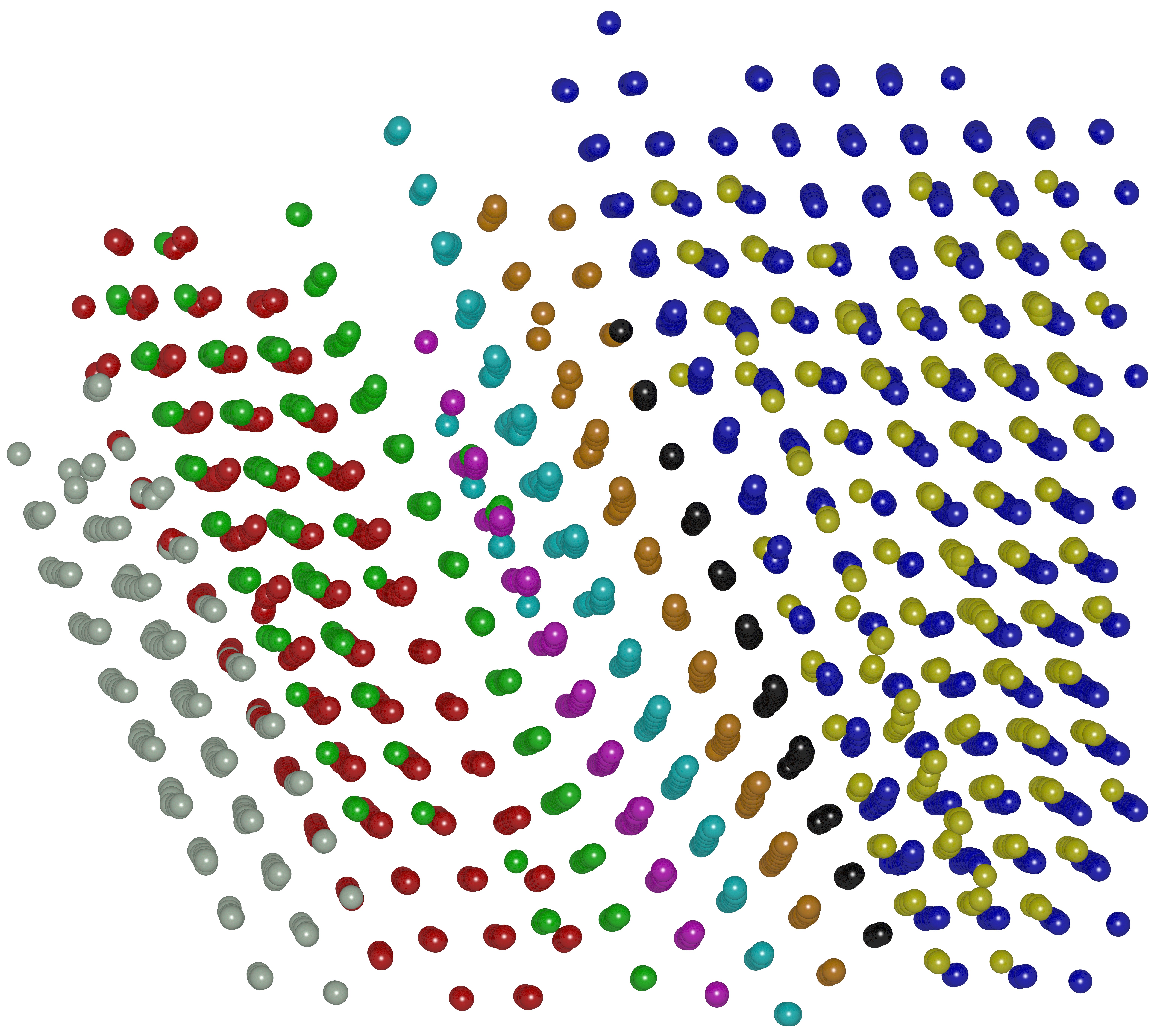}
    
    \caption[]{Analysis of experimentally self-assembled sample of gold NPs bound inside of tetrahedral frames. (A) A heat map representing  the SymBOP signal in the spherical coordinates. The hot spots allow to identify the  three orthogonal symmetry axes of the largest domain (black stars) of the domain.
    (B) \newtext{A 2D projection of the 3D domains} found from percolating by bond percolation procedure. Many domains are identified for this sample. At least two of the cubic diamond domains are identified as interpenetrating FCC lattices (blue/yellow and red/green). \newtext{ Full 3D visualization can be generated from XYZ file included in Supplementary materials.}}
    \label{fig:exp_coh_grain_bounds_ql4_plots}
\end{figure}

The second experimental example, a cubic diamond lattice, shown in Fig.~\ref{fig:exp_coh_bounds_sample}, can be seen to have interesting structures before any elaborate analysis. When the data is analyzed using SymBOPs, several domains are found. 
The global bond order parameter of the whole sample is rotated to find the orientation needed for the reference vector to coincide with the initial domains. Figure~\ref{fig:exp_coh_grain_bounds_ql4_plots}A represents the SymBOP heat map, and Fig.~\ref{fig:exp_coh_grain_bounds_ql4_plots}B shows the domains identified with our procedure. 
As one can see, there are two main domains separated by what appears to be a domain grain boundary. Each of the two domains has a cubic diamond structure. Remarkably, our analysis not only identifies  the  domains but also pulls out  pairs of interpenetrating FCC sublattices that make up the cubic diamond structure (seen in blue/yellow and red/green).    The two domains of cubic diamond are not fully aligned with each other and are separated by several additional layers, one of which is incomplete. In other words, there is a single dislocation that due to the finite thickness of the sample results in a finite angle between the two domains.  Alternatively, this can be seen as a finite-size  case  of a twin boundary as the two large domains are related by an approximate reflection symmetry with respect to the plane of the incomplete layer.


\section{\label{sec:Conclusion} Conclusion}

In this paper, we have extended the SymBOP method to analyze data that lack orientational information of the constituent particles, effectively treating the particles as isotropic for the analysis. The method described above was applied to experimental  DNA-frame-assembled NP lattices.

The key aspect of the analysis relies on finding a reference vector that serves as a template to identify bonds in the data with the given orientation and symmetry. We demonstrated the application of SymBOPs in identifying coherent domains in both simulated and experimental data. By computing all three Euler angles, the reference vector can fully match the domains present in the sample. In the examples presented, SymBOPs were employed to search for domains with cubic symmetry, specifically FCC  and cubic diamond crystals.  

In the first experimental sample, SymBOPs successfully identified two large domains with an FCC crystal structure, separated by a domain boundary. We further demonstrated the power of SymBOPs by analyzing a second experimental example of an imperfect cubic diamond  and  revealing its potential in identifying topological defects.

It is important to note that if the individual orientations of anisotropic particles are known, the method discussed in Ref.~\citenum{logan2022symmetry} should be employed instead. Pairing SymBOPs with polyhedral nematic order parameters provides a powerful tool for analysis. In this work, we have demonstrated that even without explicit orientation information, SymBOPs can be effectively utilized for data analysis by symmetrizing the bond order parameters. Further development of our approach may include its combination with  Machine Learning(ML), similar to ML-based structure classification demonstrated in ref. \cite{Boattini2019}.

Overall, the extension of SymBOPs to data without orientational information expands the scope of its application and offers valuable insights into the characterization of coherent domains. Our generic approach allows one to extend the SymBOP analysis beyond the cubic symmetry discussed in this paper.  The versatility and ease of iteration make SymBOPs a valuable tool for investigating ordering phenomena in a variety of systems and materials.

\section*{Software and data availability}
 The software created for the SymBOP analysis is publicly available from  GitHub depositary: \url{https://github.com/jalogan/SymBOPs_for_Isotropic_Particles}. In addition, the depository contains data in XYZ format that can be used to generate 3D images of the samples studied.

\begin{acknowledgments}
This research was done at and used resources of the Center for Functional Nanomaterials and the Hard X-ray Nanoprobe Beamline (HXN) at 3-ID of the National Synchrotron Light Source II, which are the US Department of Energy, Office of Science facilities at Brookhaven National Laboratory under contract DE-SC0012704. The DNA-assembly work was supported by  the US Department of Energy, Office of Basic Energy Sciences, Grant DE-SC0008772. 
\end{acknowledgments}


\appendix

\section{Experimental Methods}
The experimental data were obtained and processed by using methods previously published in Ref. \citenum{michelson2022three}. Below is a brief  overview of this methodology.

{\bf DNA Frame synthesis}
Two tetrahedral origami frames ($A$ and $B$) are synthesized ($20$nM) using M13mp18 and short staple DNA strands ($1:7$ ratio) in $1$x TAE buffer ($40$mM Tris Acetate, $1$mM EDTA) with $12.5$mM $Mg^{2+}$. The mixture was annealed slowly from $90^o$C to RT over the course of 20hrs for formation.

{\bf DNA-NP Functionalization}
Functionalization was performed on either $20nm$ or $15nm$ gold NPs for FCC (intention for NP placement at tetrahedron vertices) and Diamond (intention for NP placement on the interior of the tetrahedral frame) assemblies respectively. Gold NPs functionalized with citrate were purchased from Ted Pella. NPs were modified with oligonucleotides by adding thiol modified oligos to a solution at a mole ratio of $800:1$ (for $20nm$ NP) or $1500:1$ (for $15nm$ NP) between DNA and NPs. After mixing for 2 hours, the solution is buffered to pH $7.4$ using 10mM phosphate buffer. Salt (NaCl) was added to the solution slowly to a final concentration of 0.3M over the course of 6 hours. The mixture was lightly shaken overnight and excess reagents were removed by centrifugation of $30$ min at 10k r.c.f. four times with DI water.

{\bf Superlattice Assembly}
Functionalized NP and DNA frames (A only for FCC or A/B for Diamond) were mixed with a ratio of 1:1 and annealed from 50-RT over 72 hours at $-0.2^o$C/hour in an Eppendorf PCR Mastercycler Pro S.

\begin{figure}[htb!]
    \centering
\includegraphics[width=1\linewidth]{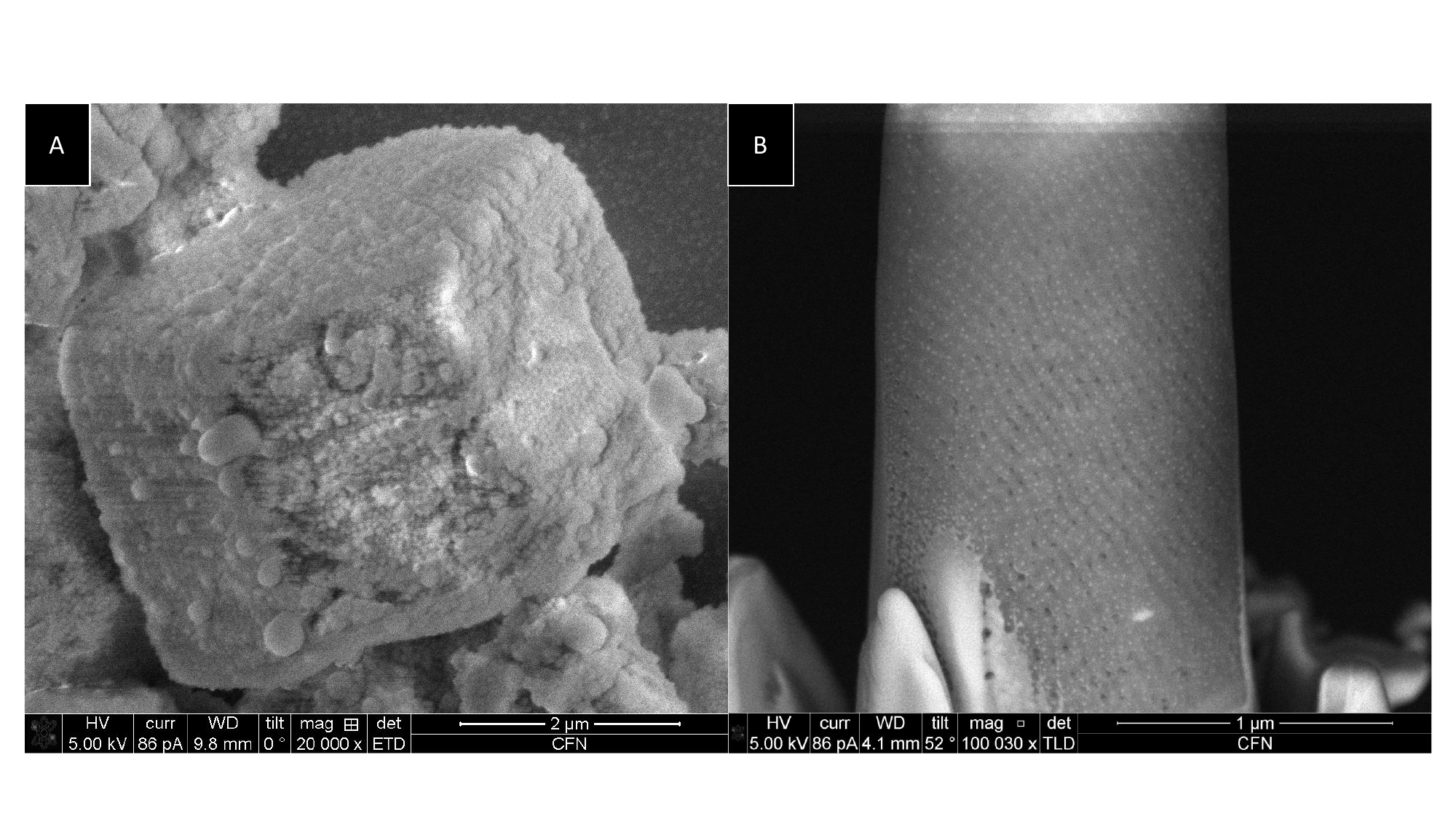}
    \caption[]{SEM of Diamond Tetrahedron Superlattice. A) SEM of silica-coated nanolattice assembly prior to sectioning. B) High-resolution SEM micrograph of a shaped superlattice for SHXM, note NPs are bright particulates on a gray mass of tetrahedral superlattice.}
   \label{FigS1}
\end{figure}

{\bf Silication}
Lattices were centrifuged and supernatant replaced with 0.1xTAE with 10mM $Mg$. Samples were either brought into a cold room at $4^o$C or RT and incubated with (3- Aminopropyl)-triethoxysilane (APTES) for 30 minutes. Tetraethoxysilane (TEOS) was then slowly added to the lattice at a vigorous mix speed over 2 hours, incubated at $4^o$C for an additional 2 hours, then slowly brought to RT over 24 hours in a thermomixer at 1000 RPM from $8-20^o$C degrees. The sample was then calcined at $550^o$C for three hours to achieve full conversion to an inorganic phase and ensure robustness for the X-ray experiment. An SEM image of a typical silicated superlattice is shown in Fig. \ref{FigS1}A.   

\begin{figure}[htb!]
    \centering
\includegraphics[width=1\linewidth]{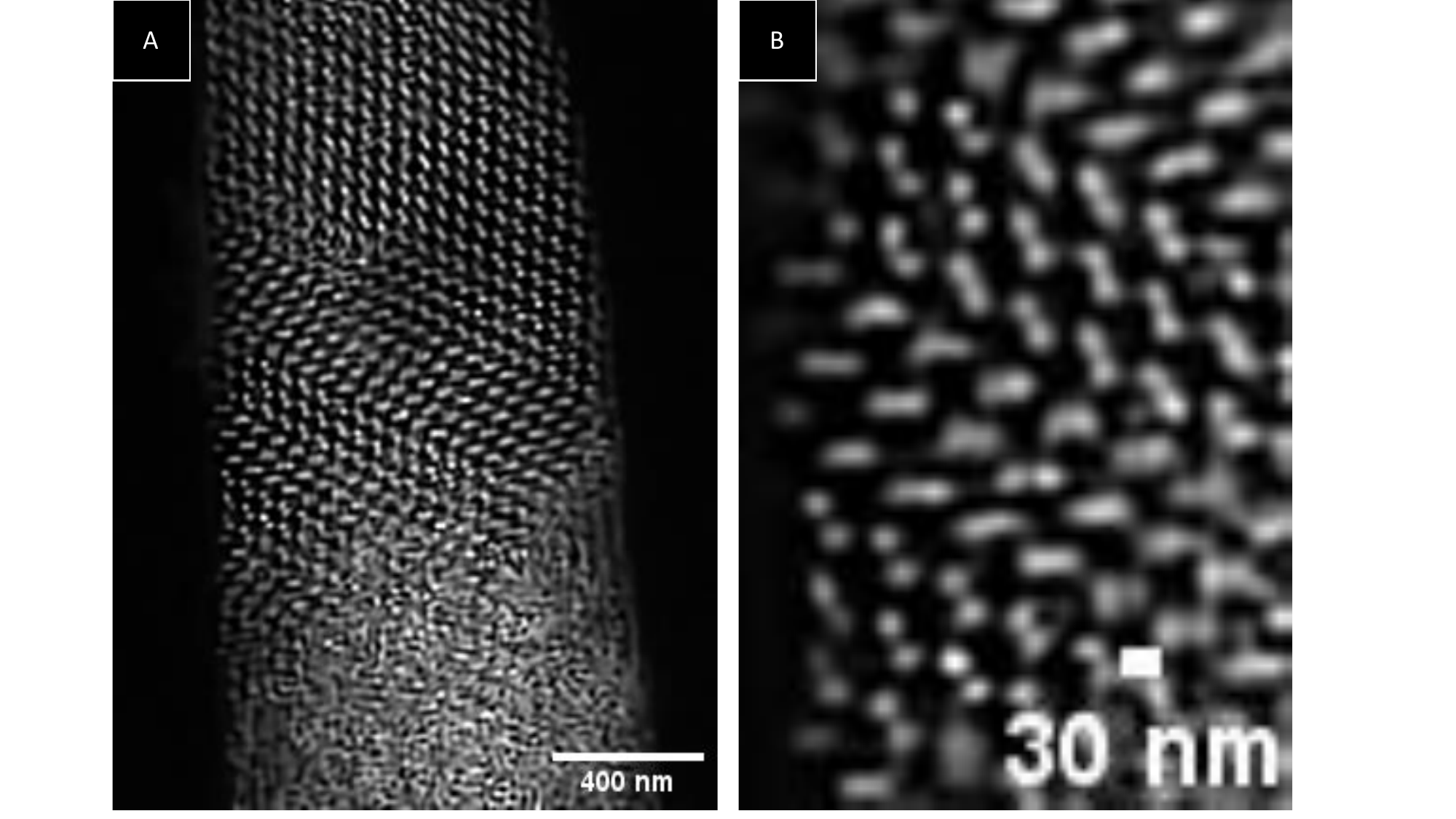}
    \caption[]{SHXM Deconvoluted Fluorescence Gold (Au) Map. A) Deconvoluted result which shows 15nm NP in a diamond nanolattice B) Zoomed in the region of the left-hand section with three coherent grain boundaries (note change in direction of dumbbell-like pattern).}
   \label{FigS2}
\end{figure}

{\bf Scanning Hard X-ray microscopy (SHXM)}
A select nanolattice was prepared using a FEI focused ion beam (Helios) to prepare a sample with proper dimensions for tomography (see Fig. \ref{FigS1}B). A monochromatic beam at 12 keV was selected and focused using MLL optics to produce a nanobeam \cite{nanofutures}. 135 projections were obtained spanning 90- to 90+ with an average of 2 degrees spacing in $\theta$. Projection images were obtained through a scan consisting of 160x200 points and a step size of 10 nm. These were aligned and reconstructed using methods outlined in \citenum{michelson2022three}, resulting in  real-space images similar to Fig. \ref{FigS2}. The final segmented Au particles were then obtained from the reconstructed data and prepared in XYZ format for subsequent analysis by SymBOP.

\bibliography{main.bib}

\pagebreak

\end{document}